\RequirePackage{lineno}

\documentclass[aps,prd,twocolumn,superscriptaddress,showpacs,floatfix,nofootinbib]{revtex4}

\usepackage{graphicx}  
\usepackage{dcolumn}   
\usepackage{bm}        
\usepackage{amssymb}   

\usepackage{lineno}
\usepackage{setspace}

\newcommand{\pbar}{\bar{p} }
\newcommand{\ppbar}{ p \bar{p} }
\newcommand{\ttbar}{ t \bar{t} }
\newcommand{\Ht}{H_T}
\newcommand{\pt}{p_T}
\newcommand{\met}{E_T\hspace{-1.1em}/\hspace{0.8em}}
\newcommand{\metb}{E_{T}\hspace{-1.1em}/\hspace{0.7em}}
\newcommand{\metc}{E_{T}\hspace{-1.1em}/\hspace{0.5em}}
\newcommand{\vmet}{\vec{E_T}\hspace{-1.1em}/\hspace{0.7em}}

\begin{document}





\title{
  Measurement of the Top Pair Production Cross Section 
  in the Dilepton Decay Channel
  in $\ppbar$ Collisions at $\sqrt{\rm s} = 1.96$~TeV
}

\affiliation{Institute of Physics, Academia Sinica, Taipei, Taiwan 11529, Republic of China} 
\affiliation{Argonne National Laboratory, Argonne, Illinois 60439, USA} 
\affiliation{University of Athens, 157 71 Athens, Greece} 
\affiliation{Institut de Fisica d'Altes Energies, Universitat Autonoma de Barcelona, E-08193, Bellaterra (Barcelona), Spain} 
\affiliation{Baylor University, Waco, Texas 76798, USA} 
\affiliation{Istituto Nazionale di Fisica Nucleare Bologna, $^{bb}$University of Bologna, I-40127 Bologna, Italy} 
\affiliation{Brandeis University, Waltham, Massachusetts 02254, USA} 
\affiliation{University of California, Davis, Davis, California 95616, USA} 
\affiliation{University of California, Los Angeles, Los Angeles, California 90024, USA} 
\affiliation{Instituto de Fisica de Cantabria, CSIC-University of Cantabria, 39005 Santander, Spain} 
\affiliation{Carnegie Mellon University, Pittsburgh, Pennsylvania 15213, USA} 
\affiliation{Enrico Fermi Institute, University of Chicago, Chicago, Illinois 60637, USA}
\affiliation{Comenius University, 842 48 Bratislava, Slovakia; Institute of Experimental Physics, 040 01 Kosice, Slovakia} 
\affiliation{Joint Institute for Nuclear Research, RU-141980 Dubna, Russia} 
\affiliation{Duke University, Durham, North Carolina 27708, USA} 
\affiliation{Fermi National Accelerator Laboratory, Batavia, Illinois 60510, USA} 
\affiliation{University of Florida, Gainesville, Florida 32611, USA} 
\affiliation{Laboratori Nazionali di Frascati, Istituto Nazionale di Fisica Nucleare, I-00044 Frascati, Italy} 
\affiliation{University of Geneva, CH-1211 Geneva 4, Switzerland} 
\affiliation{Glasgow University, Glasgow G12 8QQ, United Kingdom} 
\affiliation{Harvard University, Cambridge, Massachusetts 02138, USA} 
\affiliation{Division of High Energy Physics, Department of Physics, University of Helsinki and Helsinki Institute of Physics, FIN-00014, Helsinki, Finland} 
\affiliation{University of Illinois, Urbana, Illinois 61801, USA} 
\affiliation{The Johns Hopkins University, Baltimore, Maryland 21218, USA} 
\affiliation{Institut f\"{u}r Experimentelle Kernphysik, Karlsruhe Institute of Technology, D-76131 Karlsruhe, Germany} 
\affiliation{Center for High Energy Physics: Kyungpook National University, Daegu 702-701, Korea; Seoul National University, Seoul 151-742, Korea; Sungkyunkwan University, Suwon 440-746, Korea; Korea Institute of Science and Technology Information, Daejeon 305-806, Korea; Chonnam National University, Gwangju 500-757, Korea; Chonbuk National University, Jeonju 561-756, Korea} 
\affiliation{Ernest Orlando Lawrence Berkeley National Laboratory, Berkeley, California 94720, USA} 
\affiliation{University of Liverpool, Liverpool L69 7ZE, United Kingdom} 
\affiliation{University College London, London WC1E 6BT, United Kingdom} 
\affiliation{Centro de Investigaciones Energeticas Medioambientales y Tecnologicas, E-28040 Madrid, Spain} 
\affiliation{Massachusetts Institute of Technology, Cambridge, Massachusetts 02139, USA} 
\affiliation{Institute of Particle Physics: McGill University, Montr\'{e}al, Qu\'{e}bec, Canada H3A~2T8; Simon Fraser University, Burnaby, British Columbia, Canada V5A~1S6; University of Toronto, Toronto, Ontario, Canada M5S~1A7; and TRIUMF, Vancouver, British Columbia, Canada V6T~2A3} 
\affiliation{University of Michigan, Ann Arbor, Michigan 48109, USA} 
\affiliation{Michigan State University, East Lansing, Michigan 48824, USA}
\affiliation{Institution for Theoretical and Experimental Physics, ITEP, Moscow 117259, Russia}
\affiliation{University of New Mexico, Albuquerque, New Mexico 87131, USA} 
\affiliation{Northwestern University, Evanston, Illinois 60208, USA} 
\affiliation{The Ohio State University, Columbus, Ohio 43210, USA} 
\affiliation{Okayama University, Okayama 700-8530, Japan} 
\affiliation{Osaka City University, Osaka 588, Japan} 
\affiliation{University of Oxford, Oxford OX1 3RH, United Kingdom} 
\affiliation{Istituto Nazionale di Fisica Nucleare, Sezione di Padova-Trento, $^{cc}$University of Padova, I-35131 Padova, Italy} 
\affiliation{LPNHE, Universite Pierre et Marie Curie/IN2P3-CNRS, UMR7585, Paris, F-75252 France} 
\affiliation{University of Pennsylvania, Philadelphia, Pennsylvania 19104, USA}
\affiliation{Istituto Nazionale di Fisica Nucleare Pisa, $^{dd}$University of Pisa, $^{ee}$University of Siena and $^{ff}$Scuola Normale Superiore, I-56127 Pisa, Italy} 
\affiliation{University of Pittsburgh, Pittsburgh, Pennsylvania 15260, USA} 
\affiliation{Purdue University, West Lafayette, Indiana 47907, USA} 
\affiliation{University of Rochester, Rochester, New York 14627, USA} 
\affiliation{The Rockefeller University, New York, New York 10065, USA} 
\affiliation{Istituto Nazionale di Fisica Nucleare, Sezione di Roma 1, $^{gg}$Sapienza Universit\`{a} di Roma, I-00185 Roma, Italy} 

\affiliation{Rutgers University, Piscataway, New Jersey 08855, USA} 
\affiliation{Texas A\&M University, College Station, Texas 77843, USA} 
\affiliation{Istituto Nazionale di Fisica Nucleare Trieste/Udine, I-34100 Trieste, $^{hh}$University of Trieste/Udine, I-33100 Udine, Italy} 
\affiliation{University of Tsukuba, Tsukuba, Ibaraki 305, Japan} 
\affiliation{Tufts University, Medford, Massachusetts 02155, USA} 
\affiliation{Waseda University, Tokyo 169, Japan} 
\affiliation{Wayne State University, Detroit, Michigan 48201, USA} 
\affiliation{University of Wisconsin, Madison, Wisconsin 53706, USA} 
\affiliation{Yale University, New Haven, Connecticut 06520, USA} 
\author{T.~Aaltonen}
\affiliation{Division of High Energy Physics, Department of Physics, University of Helsinki and Helsinki Institute of Physics, FIN-00014, Helsinki, Finland}
\author{B.~\'{A}lvarez~Gonz\'{a}lez$^v$}
\affiliation{Instituto de Fisica de Cantabria, CSIC-University of Cantabria, 39005 Santander, Spain}
\author{S.~Amerio}
\affiliation{Istituto Nazionale di Fisica Nucleare, Sezione di Padova-Trento, $^{cc}$University of Padova, I-35131 Padova, Italy} 

\author{D.~Amidei}
\affiliation{University of Michigan, Ann Arbor, Michigan 48109, USA}
\author{A.~Anastassov}
\affiliation{Northwestern University, Evanston, Illinois 60208, USA}
\author{A.~Annovi}
\affiliation{Laboratori Nazionali di Frascati, Istituto Nazionale di Fisica Nucleare, I-00044 Frascati, Italy}
\author{J.~Antos}
\affiliation{Comenius University, 842 48 Bratislava, Slovakia; Institute of Experimental Physics, 040 01 Kosice, Slovakia}
\author{G.~Apollinari}
\affiliation{Fermi National Accelerator Laboratory, Batavia, Illinois 60510, USA}
\author{J.A.~Appel}
\affiliation{Fermi National Accelerator Laboratory, Batavia, Illinois 60510, USA}
\author{A.~Apresyan}
\affiliation{Purdue University, West Lafayette, Indiana 47907, USA}
% ...
\collaboration{CDF Collaboration\footnote{With visitors from $^a$University of Massachusetts Amherst, Amherst, Massachusetts 01003,
$^b$Istituto Nazionale di Fisica Nucleare, Sezione di Cagliari, 09042 Monserrato (Cagliari), Italy,
$^c$University of California Irvine, Irvine, CA  92697, 
$^d$University of California Santa Barbara, Santa Barbara, CA 93106
$^e$University of California Santa Cruz, Santa Cruz, CA  95064,
$^f$CERN,CH-1211 Geneva, Switzerland,
$^g$Cornell University, Ithaca, NY  14853, 
$^h$University of Cyprus, Nicosia CY-1678, Cyprus, 
$^i$University College Dublin, Dublin 4, Ireland,
$^j$University of Fukui, Fukui City, Fukui Prefecture, Japan 910-0017,
$^k$Universidad Iberoamericana, Mexico D.F., Mexico,
$^l$Iowa State University, Ames, IA  50011,
$^m$University of Iowa, Iowa City, IA  52242,
$^n$Kinki University, Higashi-Osaka City, Japan 577-8502,
$^o$Kansas State University, Manhattan, KS 66506,
$^p$University of Manchester, Manchester M13 9PL, England,
$^q$Queen Mary, University of London, London, E1 4NS, England,
$^r$Muons, Inc., Batavia, IL 60510,
$^s$Nagasaki Institute of Applied Science, Nagasaki, Japan, 
$^t$National Research Nuclear University, Moscow, Russia,
$^u$University of Notre Dame, Notre Dame, IN 46556,
$^v$Universidad de Oviedo, E-33007 Oviedo, Spain, 
$^w$Texas Tech University, Lubbock, TX  79609, 
$^x$IFIC(CSIC-Universitat de Valencia), 56071 Valencia, Spain,
$^y$Universidad Tecnica Federico Santa Maria, 110v Valparaiso, Chile,
$^z$University of Virginia, Charlottesville, VA  22906,
$^{aa}$Yarmouk University, Irbid 211-63, Jordan,
$^{ii}$On leave from J.~Stefan Institute, Ljubljana, Slovenia, 
}}
\noaffiliation

\date{\today}

\begin{abstract}
A measurement of the $\ttbar$ production cross section
in $\ppbar$ collisions at $\sqrt{{\rm s}}$ = 1.96 TeV using
events with two leptons, missing transverse energy, and jets
is reported. 
The data were collected with the CDF II Detector. 
The result in a data sample corresponding to an
integrated luminosity 2.8 fb$^{-1}$ is:
\begin{center}
$\sigma_{\ttbar}$ = 6.27 $\pm$ 0.73(stat) $\pm$ 0.63(syst) $\pm$ 0.39(lum) pb.
\end{center}
for an assumed top mass of 175 GeV/$c^{2}$.
\end{abstract} 


\maketitle

\section{Introduction}

This paper describes a measurement of the $\ttbar$ production cross section
in $\ppbar$ collisions
at $\sqrt{\rm s}$ = 1.96 TeV with the CDF detector at the Fermilab
Tevatron. This measurement requires the identification of both
leptons in the decay chain 
$\ttbar \rightarrow (W^+b)(W^-\overline{b})
\rightarrow (\ell^+{v}_\ell b)(\ell^-\overline{v}_\ell\overline{b})$.
 Events are selected
with two high transverse energy leptons, high missing
transverse energy ($\metc$) and at least two jets in the final
state. From the excess of events selected in the data over the
predicted background from other known standard model (SM) sources
we obtain a measurement of the production of $\ttbar$
events. 

The top quark pair production in the standard model proceeds primarily
by quark-antiquark annihilations. At the Tevatron the predictions
are 85\% quark-antiquark annihilations and 15\% gluon fusions.
At the Large Hadron Collider at $\sqrt{\rm s}$  = 14 TeV the situation
is predicted to be very different with 90\% of the production
being due to gluon-gluon fusion and 10\%  due to quark-antiquark annihilation.

This analysis improves upon a previous
measurement of the cross section using the same dilepton (DIL) selection
in a data sample with an integrated luminosity of 0.197 fb$^{-1}$~\cite{DIL}.
Unlike other CDF measurements of the $\ttbar$
cross section in the dilepton channel~\cite{ltrack}, 
where one $\ell$ is identified as $e$ or $\mu$ while the other is
identified by the presence of a high momentum central track,
the DIL analysis positively identifies both
leptons as either electrons or muons from $W$ decays or
as products of semileptonic decays of $\tau$ leptons, thus 
allowing for the comparison of 
the observed yield of $\ttbar$ decays to $ee$, $\mu\mu$ and $e\mu$ 
final states with the predictions from lepton universality.

The measurement provides a test of the QCD calculations of the
$\ttbar$ cross section~\cite{theory} in a channel which is
independent and complementary to other measurements of the $\ttbar$
cross section in higher statistics final states
where at least one $W$ boson from the top quark is reconstructed
via its hadronic decay, $W\rightarrow { q}{q^{\prime}}$. 
The dilepton final state suffers from a lower statistical precision,
as the product of the branching ratios of the semileptonic $W$ decay
BR($W^{+} \rightarrow \ell^{+} \nu$)$\times$Br($W^{-} \rightarrow \ell^{-} \nu$) 
$\approx$ 5\% with $\ell$ = {\it e} or $\mu$,
but it has a signal to background
ratio well above unity even before requiring the identification 
of one of the jets originating from a {\it b} quark. 
This analysis does not require jets in the events to have secondary vertexes 
consistent with the presence of a $b$-hadron decay as this selection would 
further reduce the acceptance by almost 50\%.

In Sec.~\ref{sec:Det} we give a short description of the detector. 
In Sec.~\ref{sec:Data} the data sample and event selection are presented. 
Section~\ref{sec:Acc} presents
the formula used for the cross section calculation and the  
measurement of the $\ttbar$ acceptance in dilepton events.  
Section~\ref{sec:Bak} describes the calculation of the backgrounds.
Systematics uncertainties are covered in Sec.~\ref{sec:Sys}. 
In Sec.~\ref{sec:control-samples} observations are
compared to predictions in control samples
characterized by the presence of two leptons plus 
high $\met$ in the final state. 
We conclude by presenting the result of our measurement
in Sec.~\ref{sec:Results}, followed by a short summary in 
Sec.~\ref{sec:Concl}.


\section{\label{sec:Det}Detector}
CDF II is  a general-purpose detector that is described in detail
elsewhere~\cite{CDF}. The components relevant to this analysis 
are briefly described here. The detector has an approximate full
angular coverage with a charged particle tracker inside a magnetic solenoid,
backed by calorimeters and muon detectors.
CDF uses a cylindrical coordinate system in which $\theta$ 
is the polar angle about an axis defined by
the proton beam and $\phi$ is the azimuthal angle 
about the beam axis. Particle pseudorapidity is defined as
$\eta=-$ln tan($\theta$/2). 

The charged-particle tracking system surrounds the beam pipe
and consists of multiple layers of silicon micro-strip detectors,
which cover a pseudorapidity region $|\eta|< 2$,
and a 3.1 m long open-cell drift chamber covering the pseudorapidity region
$|\eta|< 1$ \cite{svx,cot}.
The tracking  system  is located inside a superconducting solenoid,
which in turn is surrounded by calorimeters. 
The magnetic field has a strength of 1.4 T
and is aligned co-axially with the $p$ and $\pbar$ beams.

\par
The calorimeter system~\cite{cal} is split radially into electromagnetic (EM)
and hadronic (HAD) sections segmented in projective tower geometry, and covers  
the pseudorapidity range $|\eta|< 3.6$. 
The electromagnetic sampling calorimeters are constructed of alternating
layers of lead absorber and scintillator
whereas the hadronic calorimeters use iron absorbers.
Proportional chambers (CES) are embedded in the central electromagnetic
calorimeter at a depth of about 6X$_0$(radiation length), which is the region of maximum
shower intensity for electrons. In the plug region stereo layers of scintillator bars are placed at shower maximum.

\par
 A set of drift chambers located outside the central
calorimeters (CMU), complemented by another set behind a 60~cm iron 
shield (CMP), provides muon coverage for 
$|\eta|$ $\leq$ 0.6.
Additional drift chambers and scintillation counters (CMX) 
detect muons in the region
0.6 $\leq$ $|\eta|$ $\leq$ 1.0~\cite{muon}.

\par
Multi-cell gas Cerenkov counters~\cite{Cherenkov} 
located in the 3.7 $<$ $|\eta|$ $ <$ 4.7 region
measure the average number of inelastic $\ppbar$ collisions per 
bunch crossing and thus determine the beam luminosity. 
The total uncertainty on the luminosity
is estimated to be 5.9\%, of which 4.4\% comes 
from the acceptance and 
operation of the luminosity monitor
and 4.0\% from the uncertainty in the inelastic $\ppbar$ cross section.

\section{\label{sec:Data}Data Sample and Event Selection}
This analysis is based on an integrated luminosity of 2.8~fb$^{-1}$ collected
with the CDF II detector between March 2002 and April 2008.  The
data are collected with an inclusive lepton trigger that requires an
electron (muon) with $E_{T}>$ 18~GeV  ($p_{ T}>$ 18~GeV/$c$). 
The transverse energy and transverse momentum are defined 
as $E_T = E\,\sin\theta$ and $p_T = p\,\sin\theta$, 
where $E$ is energy measured in the calorimeter and $p$
is momentum measured by the tracking system. 
From this inclusive lepton data set, events with a
reconstructed isolated electron of $E_T$ (muon of $p_T$)
greater than 20 GeV(GeV/$c$) are selected.
Isolation is defined as the calorimeter energy deposited in a cone of 
radius $\Delta R \equiv\sqrt{(\Delta\eta)^2+(\Delta\phi)^2}=0.4$ in 
$\eta-\phi$ space centered around the lepton, 
minus the energy deposited by the lepton itself.
Details on electron and muon identification, or lepton ID, criteria
used in this analysis are contained in reference~\cite{leptonID}. 
Electrons are identified by matching clusters of localized energy deposition
in the calorimeter to tracks 
reconstructed using hits from the tracking chambers and
the extra constraint provided by the position of the
beam line in the transverse direction. We further require that
the energy deposition in the electromagnetic section of the
calorimeter exceeds the energy measured in
the corresponding hadronic section and the lateral cluster energy
profile agrees with shapes derived from electron beam-test data.
Muons are identified by matching tracks to
minimum ionizing-like clusters in the calorimeter and to stubs,
or sets of radially aligned hits, in the muon chambers.
Leptons passing all of the lepton ID cut and with isolation less 
than 10\% of the lepton energy are defined 
as ``tight''. They can be of one of four categories: 
electrons reconstructed in the central (CEM) or 
plug (PHX) calorimeter and muons pointing to the regions covered by both
layers of 
the two central muon chambers (CMUP) or by the muon extension chambers (CMX).
These tight leptons are also required to be the objects that trigger the 
event, with the exception of PHX electrons which are allowed in events 
triggered by non isolated CEM, CMUP or CMX.

\par
A second electron of $E_T$ (muon of $p_T$) 
greater than 20 GeV(GeV/$c$) is also required
using looser identification cuts and without the isolation requirement.
``Loose'' leptons are either electrons or muons 
which pass the same identification cuts as the tight leptons, 
but fail the isolation requirement. Another category of loose
leptons has no tight lepton counterpart and is made of
muons with tracks pointing to regions covered by only 
one of the two central muon chambers (CMU, CMP) or with
tracks of energy deposition corresponding to a minimum ionizing particle
and pointing to regions not covered by a muon chamber (CMIO). 
CMIO muons, as well as PHX electrons, must be isolated.

\par
Each dilepton candidate must contain at least one tight lepton and at most
one loose lepton. These requirements result in 18 different DIL dilepton 
categories, as illustrated in Sec.~\ref{sec:Den}, 
where background estimates and acceptances are calculated separately for 
each category.
Events with more than two tight or loose leptons in the final state 
are rejected
as they come mostly from background sources like $WZ$ and $ZZ$ events.
Another source of trilepton background comes from
Drell-Yan $Z/\gamma^* \to$ $e^+e^-$ events with a radiated photon
converting into an asymmetric $e^+e^-$ pair, that is a conversion pair
where one electron does not reach the minimum track $p_T$ threshold 
of 500 MeV/$c$ needed for the electron to not be trapped inside
the tracking chamber.
The loss in signal efficiency from removal of events with three or more
leptons is only 0.4\%.

\par
A fraction of events passing the dilepton selection does not originate
from $\ppbar$ collisions but from beam interactions with the
detector and shielding material or from cosmic ray sources.
These events are removed by requiring 
a reconstructed vertex consistent with originating 
from the beam interaction region and within 60~cm of the center of the
detector along the z-direction. We also require that the timing of tracks 
in dimuon events be consistent with both muons traveling from the center of the
detector outward into the tracking chamber~\cite{FemFch_Run1}.
Electrons from conversions of photons in the detector material
are removed by identifying events with a 
track near the electron track of opposite curvature and consistent
with coming from a $\gamma \rightarrow$ $e^+e^-$ vertex. 

\par
Jets are reconstructed from the calorimeter towers
using a cone algorithm with fixed radius 
$\Delta{\rm R}$ = 0.4 in $\eta-\phi$ space~\cite{jetalg}.
The jet $E_T$ is corrected for detector effects due 
to calorimeter dead zones and to  non-linear tower response to
deposited energy. 
These effects are convoluted to provide
 the jet energy scale (JES) correction factor
which estimates the energy of the originating parton 
from the measured energy of a jet~\cite{jetcorr}. 
Jets used in the DIL selection are required to have 
corrected $E_T > 15$~GeV and $|\eta| < 2.5$.

\par 
We further impose two cuts based on the kinematical properties of 
the event:
the first is a cut on the missing transverse energy $\metc$~\footnote{
 The scalar quantity $\metc$ is the magnitude of the 
missing transverse energy vector $\vmet$ defined as 
the opposite of the sum over the calorimeter towers 
$\sum_i {E}^i_{T}{\bm n}_i$ of the 
transverse energy measured in each tower, E$^i_{T}$,
times the unit vector in the azimuthal plane that points 
from the beam line to the $i^{th}$ calorimeter tower, ${\bm n}_i$.}
which measures the transverse energy of the neutrinos via
the imbalance of the energy detected in the calorimeter, after
correcting for the presence of muons.
We require that $\metc >$ 25~GeV,
which is strengthened to a $>$ 50~GeV requirement 
if any lepton or jet is closer than
20$\rm ^o$ to the $\met$ direction. This cut, in the following
referred to as L-cut, is used to reject mainly 
$Z \rightarrow$ $\tau^+\tau^-$ and events
with mismeasured $\met$ generated by jets pointing to cracks in the calorimeter.
The second, or $Z$-veto cut, aims at reducing the contamination of 
dilepton decays of the $Z$ boson by
requiring high missing $E_T$ significance 
for $ee$ and $\mu\mu$ events with
dilepton invariant mass in 76 to 106 GeV/$c^2$ region.  
Missing $E_T$ significance, or MetSig, is defined 
as $\metc /$$\sqrt{{ E}^{\rm sum}_{ T}}$, where
E$^{ sum}_{ T}$ is the sum of transverse energies deposited in all 
calorimeter towers. This variable separates events with real
$\metb$ due to neutrinos from events where the $\metb$
is due to energy measurement fluctations or 
energy loss in calorimeter cracks. These second category
of events is expected to have a degraded $\metb$ resolution.
In the DIL selection, we use a cut of MetSig $>$ 4~{GeV}$^{(1/2)}$.
\par
Events in the DIL dilepton sample passing the L-cut
and $Z$-veto cut become $\ttbar$ candidate events if 
they have at least 2 jets,
if the two leptons are of opposite charge (OS), 
and if $\Ht$ the transverse energy sum of leptons, neutrinos
and jets is greater than 200~GeV.
Events in the DIL dilepton sample reconstructed with 0 or 1
jet are used as control samples for the
background estimation.

\section{\label{sec:Acc}Cross Section}
In this analysis we measure the cross section by using the formula
\begin{equation}
  \label{eq:Xsec}
  { \sigma_{\ttbar} = \frac{{\rm N}_{\rm obs}-{\rm N}_{\rm bkg}}
                            {\mathcal{A} \times \mathcal{L}}}
 \end{equation}
where N$_{\rm obs}$ is the number of dilepton candidate events,
and N$_{\rm bkg}$ is the total background. The denominator
is the product for the acceptance for $\ttbar$ candidate events, $\mathcal{A}$,
and of the data set luminosity $\mathcal{L}$.

\par
The acceptance, which in our definition
includes the effects of the detector geometrical acceptance, lepton
identification and $\ttbar$ to dilepton selection, is measured using the
{\sc pythia} Monte Carlo program~\cite{pythia} to simulate $\ttbar$ events 
of all three decay modes (hadronic, lepton + jets, and dilepton)
with an assumed M$_{\rm top}$ = 175~GeV$/{c}^2$. 
Monte Carlo simulated events are required to have both $W$ bosons from top
quarks decaying to a lepton plus neutrino, where the lepton can be 
either an electron or muon.
The $\ttbar$ Monte Carlo simulation acceptance prediction 
of 0.808 $\pm$ 0.004 (stat)\% is corrected by taking
into account differences observed between data and the Monte Carlo simulation
modeling of the detector response in independent control samples.
The following sections describe the implementation and checks on
the acceptance correction procedure.

\subsection{\label{sec:Den}Signal Acceptance}
The available statistics for each sub sample of DIL events can be maximized by
requiring that only detector parts essential for the identification of
a particular lepton category be fully functioning.
For example, PHX electrons require hits in the silicon inner
vertex detector to reconstruct their tracks. 
Hence the identification of events with PHX electrons is limited
to data taken with ``good'', i.e. fully functioning, silicon detectors but no
such requirements is imposed on events where the electron is central.
To accommodate this approach, we rewrite the
denominator of the cross section formula in equation~\ref{eq:Xsec}
as the sum of the acceptance for each DIL dilepton category
$\mathcal{A}_i$, weighted by the luminosity relative to that
category $\mathcal{L}_i$: 
\begin{equation}
  {\cal A} \times {\cal L}  =  \sum_{i} {\cal A}_i \times {\cal L}_{i}
 \label{eq:Xsec_i}
\end{equation}
In this analysis four different luminosity samples are used, 
corresponding to the integrated luminosity of runs with
fully functional sub-detectors for trigger CEM electrons and CMUP muon, 
either ignoring the status of the silicon detectors (2826~pb$^{-1}$) 
or requiring ``good'' silicon (2676~pb$^{-1}$),
and runs fully functional also for trigger CMX muons, either ignoring 
(2760~pb$^{-1}$) or requiring (2623~pb$^{-1}$) ``good'' silicon.
In defining runs good for CEM, CMUP or CMX leptons, the distinction between
isolated and non-isolated leptons is irrelevant.

\par
The acceptances $\mathcal{A}_i$ can be factorized in terms
of the two leptons $\ell_1$ and $\ell_2$ comprising the DIL category $i$ 
according to the following formula
\begin{equation}
  {\cal A}_{i}  =  {\cal A}_{\ell_1\ell_2} \times {\rm C}_{\ell_1\ell_2} 
 \label{eq:A_i}
\end{equation}
where the ${\cal A}_{\ell_1\ell_2}$
are the raw {\sc pythia} $\ttbar$ MC efficiencies for events with 
reconstructed leptons $\ell_1$ and $\ell_2$ passing the full DIL selection and
the C$_{\ell_1\ell_2}$  are correction factors specific for that lepton pair.
The correction factors are in turn calculated as:
\begin{equation}
  {\rm C}_{\ell_1\ell_2} = \epsilon_{z_0} \times 
  ( \epsilon_{trg_1} + \epsilon_{trg_2} - \epsilon_{trg_1}\epsilon_{trg_2} )
   \times  {\rm SF}_{\ell_1} {\rm SF}_{\ell_2}
 \label{eq:C12}
\end{equation}
where $\epsilon_{z_0}$ is an event efficiency while
$\epsilon_{trg_i}$ and ${\rm SF}_{\ell_i}$ are
single lepton trigger efficiency and 
identification efficiency scale factors, respectively. 
The factor $\epsilon_{z_0}$ accounts for the efficiency of
$\rm \pm$ 60~cm cut on the z-position of the reconstructed event vertex.
By using a sample of minimally biased inelastic interactions, we find that
this cut accepts 96.63 $\pm$ 0.04 (stat)\% of the full CDF luminous region.
The lepton trigger efficiencies $\epsilon_{trg_i}$ are measured 
in data samples selected with independent sets of triggers and
found to be around 90\% or better. 
Finally, the scale factors ${\rm SF}_{\ell_i}$ are calculated as ratios
of lepton identification efficiencies measured in data
and in Monte Carlo simulations.

\par
Table~\ref{tab:lep_cor} lists all the factors used in the 
acceptance correction. Their central values
are the luminosity weighted averages over different data taking periods, and
the quoted uncertainties are only statistical.
Using these as inputs to Eq. (\ref{eq:C12}), we obtain
values ranging from 73\% to 93\% for the
correction factors C$_{\ell_1\ell_2}$, as shown in 
Table~\ref{tab:ttop75_table},
and a total denominator (\ref{eq:Xsec}) for 
the 2.8 fb$^{-1}$  DIL cross section of 19.43 $\pm$ 0.10 pb$^{-1}$, 
where the uncertainty
comes solely from the propagation of the statistical uncertainties
of each term in Eq.~(\ref{eq:A_i}) and Eq.~(\ref{eq:C12}). 

\begin{table*}
 \caption{ Event vertex reconstruction efficiency $\epsilon_{z_0}$ and
           list, by lepton type, of trigger efficiency $\epsilon_{trg}$,
           and lepton identification efficiency scale factors SF
           defined in Eq.~(\ref{eq:C12}).
           These are the luminosity weighted averages of 
           efficiencies calculated for
           different data taking periods. }
\begin{center}
   \begin{tabular}{l c}
   \hline\hline
        & \multicolumn{1}{c}
{ Vertex Reconstruction efficiency, $\epsilon_{z_0}$} \\
  \hline
        & 0.9663 $\pm$0.0004         \\
        &                            \\
   \hline
 Lepton Type  & \multicolumn{1}{c}{ Trigger efficiency, $\epsilon_{trg}$} \\
   \hline
CEM     & 0.965$\pm$0.001     \\
CMUP    & 0.917$\pm$0.002     \\
CMX     & 0.896$\pm$0.002     \\
        &                              \\
   \hline
      & \multicolumn{1}{c}{ Lepton Identification Scale Factor, SF} \\
   \hline
CEM        & 0.987 $\pm$ 0.001         \\
PHX        & 0.932 $\pm$ 0.002          \\
CMUP       & 0.927 $\pm$ 0.002          \\
CMX        & 0.973 $\pm$ 0.002          \\
CMU        & 0.97  $\pm$ 0.01          \\
CMP        & 0.90  $\pm$ 0.01          \\
CMIO       & 0.99  $\pm$ 0.01         \\
\hline\hline
\end{tabular}
\end{center}
 \label{tab:lep_cor}
\end{table*}

\begin{table*}
 \caption{ List, by dilepton category, of raw acceptance 
           ${\cal A}_{\ell_1\ell_2}$, correction factor
           $C_{\ell_1\ell_2}$ and luminosity ${\cal L}_{i}$ used in  
           the calculation of the denominator for the 2.8~fb$^{-1}$ 
           DIL cross section measurement.
           The acceptance of each category includes contributions 
           from non isolated loose leptons. 
           The ${\cal A}_{\ell_1\ell_2}$ uncertainty comes only
           from the MC statistics.
           The error in the $C_{\ell_1\ell_2}$ comes from the propagation 
	   of the dilepton efficiency uncertainties 
           of Table~\ref{tab:lep_cor}.} 
 \begin{center}
   \begin{tabular}{l c c c}
   \hline\hline
 DIL Category   &  ${\cal A}_{\ell_1\ell_2}$(\%) & C$_{\ell_1\ell_2}$ & ${\cal L}_{i}$(pb$^{-1}$)\\
   \hline 
CEM-CEM      &  0.1224 $\pm$ 0.0017 & 0.9338 $\pm$ 0.0019  & 2826 \\
CEM-PHX      &  0.0470 $\pm$ 0.0010 & 0.8658 $\pm$ 0.0027  & 2676 \\
             &                      &                      &      \\
CMUP-CMUP    &  0.0498 $\pm$ 0.0011 & 0.8189 $\pm$ 0.0025  & 2826 \\
CMUP-CMU     &  0.0191 $\pm$ 0.0007 & 0.7920 $\pm$ 0.0040  & 2826 \\
CMUP-CMP     &  0.0267 $\pm$ 0.0008 & 0.7299 $\pm$ 0.0035  & 2826 \\
CMUP-CMX     &  0.0474 $\pm$ 0.0010 & 0.8569 $\pm$ 0.0020  & 2760 \\
CMUP-CMIO    &  0.0234 $\pm$ 0.0007 & 0.8050 $\pm$ 0.0040  & 2826 \\
CMX-CMX      &  0.0106 $\pm$ 0.0005 & 0.8996 $\pm$ 0.0027  & 2760 \\
CMX-CMU      &  0.0075 $\pm$ 0.0004 & 0.8134 $\pm$ 0.0043  & 2760 \\
CMX-CMP      &  0.0115 $\pm$ 0.0005 & 0.7495 $\pm$ 0.0037  & 2760 \\
CMX-CMIO     &  0.0101 $\pm$ 0.0005 & 0.8267 $\pm$ 0.0043  & 2760 \\
             &                      &                      &      \\
CEM-CMUP     &  0.1769 $\pm$ 0.0020 & 0.8737 $\pm$ 0.0018  & 2826 \\
CEM-CMU      &  0.0349 $\pm$ 0.0009 & 0.8879 $\pm$ 0.0040  & 2826 \\
CEM-CMP      &  0.0475 $\pm$ 0.0010 & 0.8182 $\pm$ 0.0035  & 2826 \\
CEM-CMX      &  0.0845 $\pm$ 0.0014 & 0.9171 $\pm$ 0.0019  & 2760 \\
CEM-CMIO     &  0.0410 $\pm$ 0.0010 & 0.9025 $\pm$ 0.0041  & 2826 \\
PHX-CMUP     &  0.0327 $\pm$ 0.0009 & 0.7723 $\pm$ 0.0029  & 2676 \\
PHX-CMX      &  0.0147 $\pm$ 0.0006 & 0.7922 $\pm$ 0.0032  & 2623 \\
\hline\hline
   \end{tabular}
  \end{center}
 \label{tab:ttop75_table}
\end{table*}

\subsection{Check of Acceptance Corrections}
\label{dil_Zxsec}

As a cross-check of our acceptance correction procedure,
we calculate the cross section for $Z$ production for each
dielectron and dimuon category used in the DIL selection.
With this check we verify
the consistency of the correction procedure across
the different dilepton categories.

\par
We select events with $ee$ or $\mu\mu$ in the final state
and require the two leptons to have opposite charges and invariant mass
in the range 76 GeV/$c^2$ to 106 GeV/$c^2$. We follow the same 
lepton pairing used for the DIL dilepton selection.
Cosmic ray induced events and events with an identified conversion are removed
following the same criteria used for the $\ttbar$ dilepton selection.
The number of events selected by these cuts is shown in Table~\ref{tab:Zxsec}.

\par
We use $Z$/$\gamma^* \to$ $ee$ and $Z$/$\gamma^* \to \mu\mu$ 
{\sc pythia} Monte Carlo simulated samples
to calculate the raw acceptance of the
selection described above in the invariant mass 
76 GeV/$c^2 < M_{\ell \ell} < $ 106 GeV/$c^2$. 
We use a formulation for the $Z$ cross section calculation analogous 
to the $\ttbar$ cross section formulation in Eq.~(\ref{eq:Xsec}).
In particular we employ the same factorization for the
denominator correction calculation prescribed in the previous section by
Eqs.~(\ref{eq:Xsec_i} - \ref{eq:C12}).

\par 
We perform two checks: time independence of the acceptance 
correction factor for each DIL category and consistency of the 
correction procedure among different categories.
With these checks we are not trying to measure the 
$Z$ cross section but rather to determine if our understanding 
of the acceptance is correct.
The $Z$ cross section has been independently measured by CDF as
$\sigma_{Z} = 256 \pm 16$~pb~\cite{Zcross}.

\par
\begin{figure*}[htbp]
 \begin{center}
\includegraphics[width=0.8\textwidth, clip]{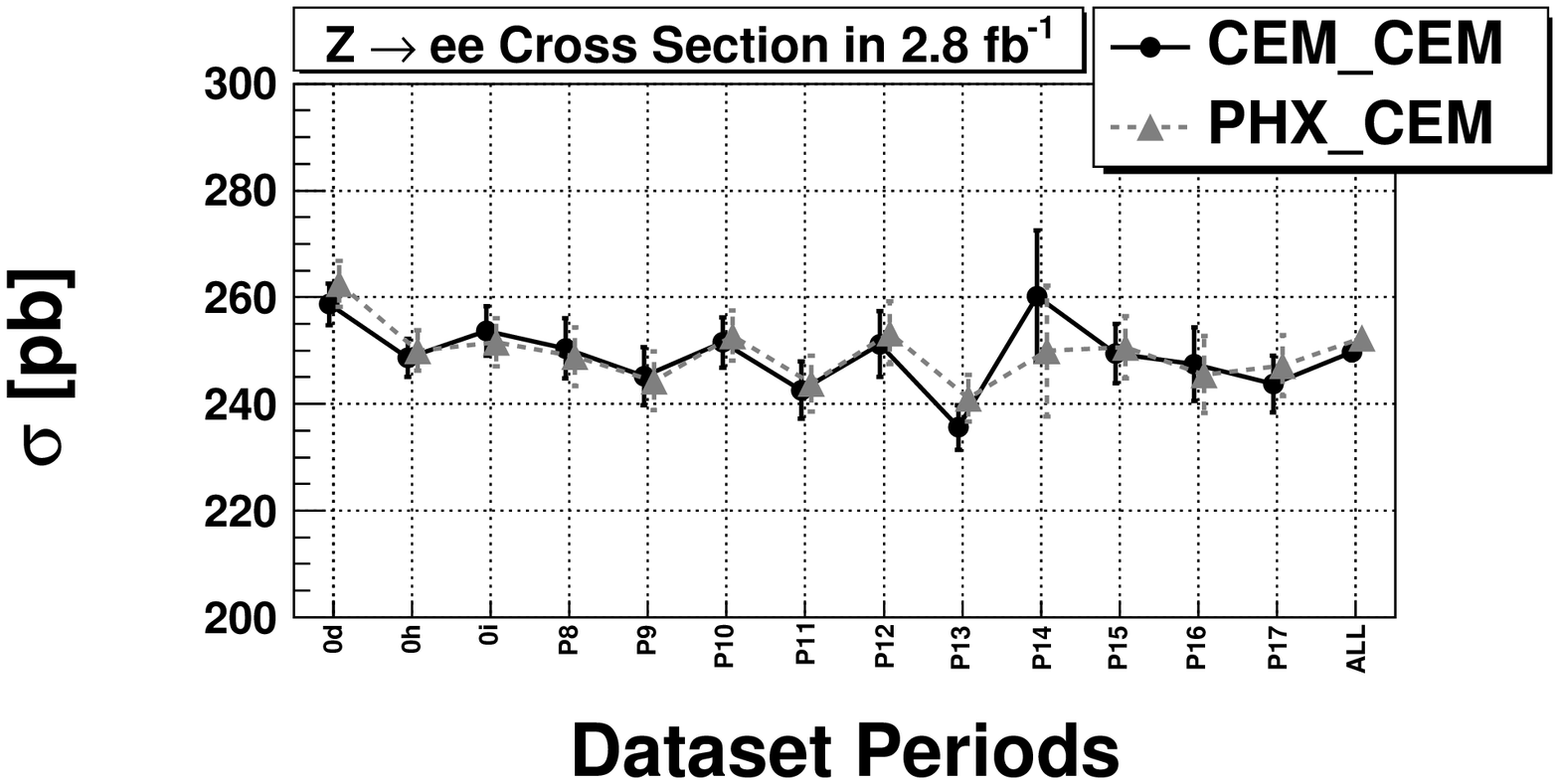}
\caption{$Z$ cross section using dilepton categories with two electrons
         as a function of data taking period. }
  \label{figZ2ee}
 \end{center}
\end{figure*}
\begin{figure*}[htbp]
 \begin{center}
\includegraphics[width=0.8\textwidth, clip]{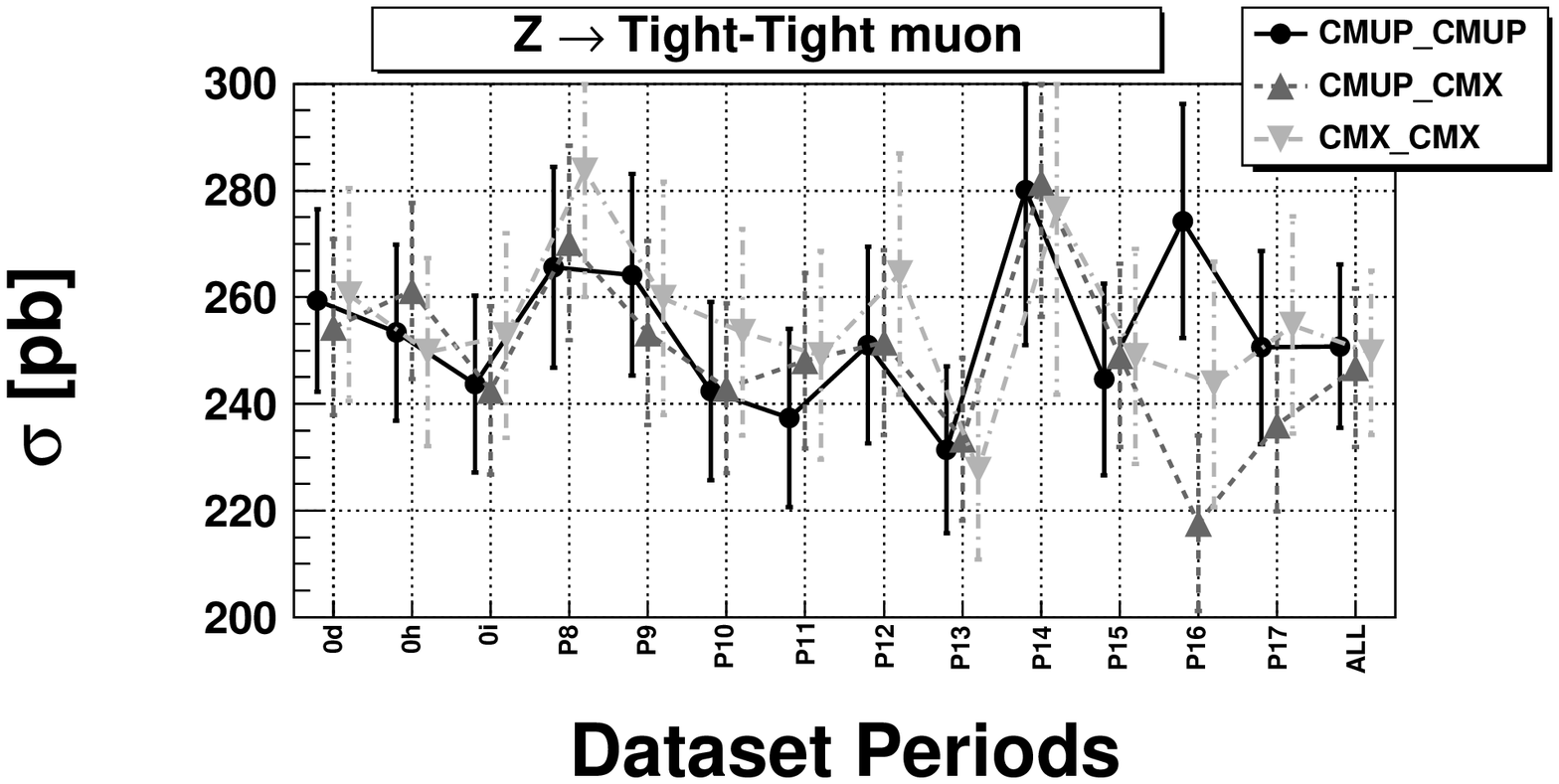}
\caption{$Z$ cross section using dilepton categories with tight muons
         as a function of data taking period.}
  \label{figZ2cmup}
 \end{center}
\end{figure*}

For the first check, we look for possible time variations
of the measured $Z$ cross section 
in different data taking periods corresponding to integrated
luminosities between 200 and 500 pb$^{-1}$.
Figure~\ref{figZ2ee} shows the result of these checks 
for dielectron channels.
The error bar in the figure reflects uncertainties of data
statistics, Monte Carlo statistics and the acceptance correction procedure.
We do not observe any systematic trend in the time
dependence of the cross section for dielectron categories.
Same conclusions hold for the dimuons channels. As an example,
Fig.~\ref{figZ2cmup} shows where the $Z$ cross section time dependence
for events with two tight muons.

%

\par
\begin{figure*}[htbp]
 \begin{center}
\includegraphics[width=0.80\textwidth, clip]{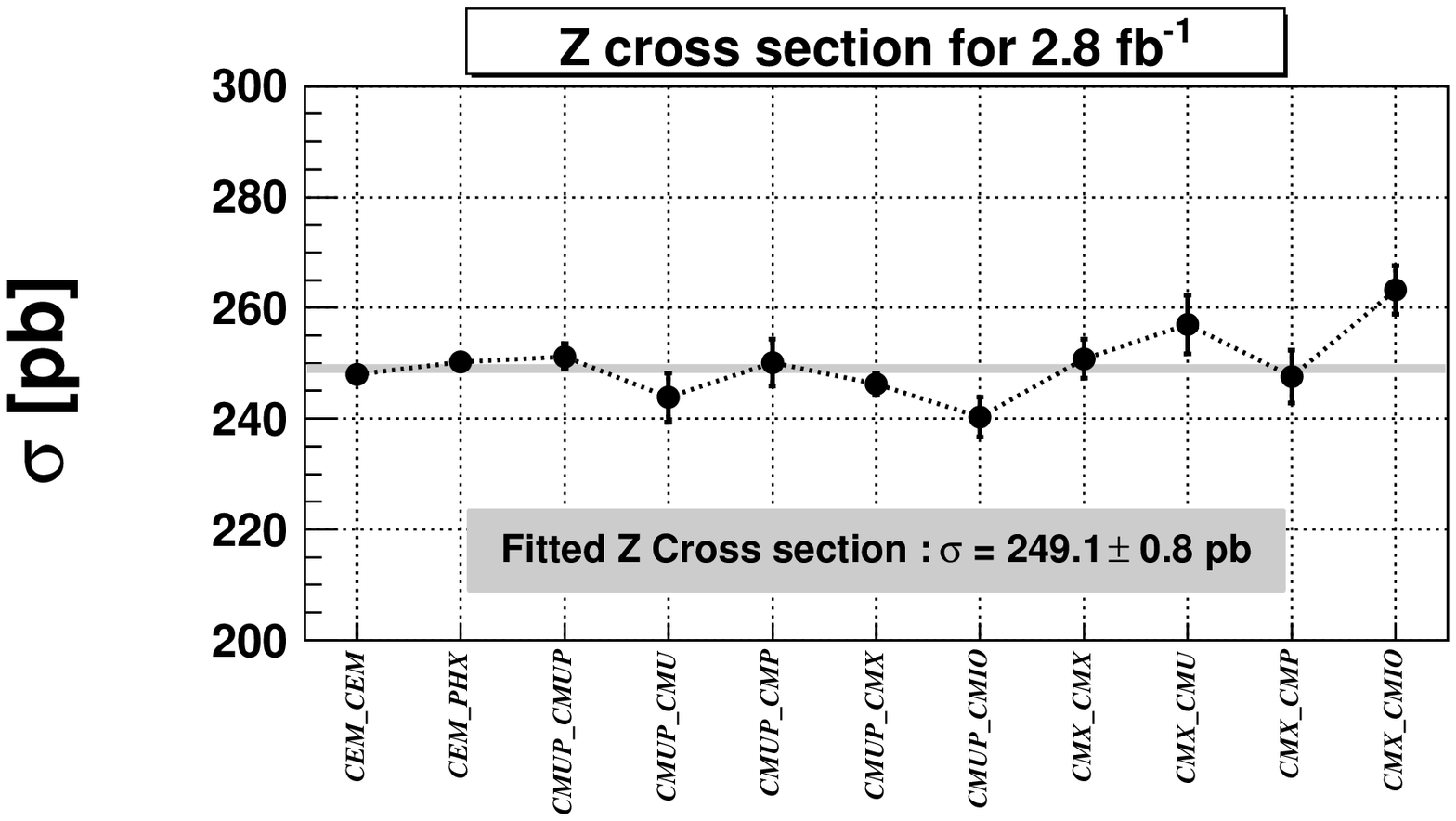}
\caption{
$Z$ peak cross section measured for each inclusive same flavor 
dilepton category using the full 2.8~fb$^{-1}$ data sample after
applying the loose muon scale factor. Not included
is a 6\% uncertainty in the luminosity measurement common to all channels.
The band represents the mean value of 249.1 $\pm$ 0.8 pb.}
  \label{figZall_data}
 \end{center}
\end{figure*}

For the second check, Table~\ref{tab:Zxsec} reports 
the cross sections measured over the whole 2.8~fb$^{-1}$ data
sample for each dilepton category, using the efficiencies and scale factors
of Table~\ref{tab:lep_cor}.
The categories with two tight leptons (CEM-CEM, CEM-PHX, CMUP-CMUP, 
CMUP-CMX and CMX-CMX), which are also the ones with the largest acceptance,
are consistent with the
theoretical prediction of 251.6$^{+2.8}_{-3.1}$ pb~\cite{Zcross}.
Categories with a loose muon (CMP, CMU or CMIO) paired to a tight muon 
show some residual variation around the average value 
which is not consistent with statistical fluctuations. 
In order to find a consistent normalization for all data, 
we perform a fit to the $Z$ cross sections in the different dilepton 
categories with three free parameters,
corresponding to a multiplicative factor in front of the selection efficiency 
of each of the three loose muon categories.
The fit returns the cross sections reported in the last column of
Table~\ref{tab:Zxsec} and an average $Z$ peak cross section
of 249.1 $\pm$ 0.8 pb, as shown in Fig.~\ref{figZall_data}. 
The $Z$ peak cross section measured here does not include the 6\%
uncertainty in the luminosity which is common to all channels and
which is the dominant uncertainty for the $\sigma_{\rm Z}$ 
measured in~\cite{Zcross}.
The free parameters returned by the fit are:
SF$_{\rm CMP}^{Z} = 0.977 \pm 0.011$,
SF$_{\rm CMU}^{Z} = 1.072 \pm 0.011$ and SF$_{\rm CMIO}^{Z} = 0.954 \pm 0.010$.
They are folded into the acceptance correction procedure as
additional scale factors to be multiplied 
by the lepton identification scale factors of 
Table~\ref{tab:lep_cor} for the appropriate categories.
After the fit, the single $Z$ cross section measurements are consistent with
each other within uncertainties, with the possible exception of
categories containing one CMIO loose muon for which we observe a maximum
deviation equal to 10\% of the average value.
This systematic deviation affects only 10\% of the DIL $\ttbar$ raw acceptance,
corresponding to the summed contributions of any dilepton pair
containing a CMIO muon in Table~\ref{tab:ttop75_table}. 
Therefore we estimate a final 1\% systematic uncertainty on the
acceptance due to the correction procedure.

\begin{table*}
 \caption{ Number of selected events and $Z$ cross section 
           in 2.8~fb$^{-1}$ for the different $ee$ and $\mu\mu$ 
           dilepton categories with dilepton invariant mass in the range 
           76 GeV/$c^2$ to 106 GeV/$c^2$. Results are given both for the
           original and the fitted $Z$ cross sections.
           The cross section uncertainties
           are only from the data statistics and from the propagation of the
           uncertainty in the single lepton efficiency 
           of Table~\ref{tab:lep_cor}.}
 \begin{center}
   \begin{tabular}{l c c c}
 \hline\hline
 Category   &  Number of Events &  $Z$ Cross Section (pb) Original & Fitted \\
   \hline     	   
CEM-CEM       & 50519   & 248.0 $\pm$ 1.4 & 248.0 $\pm$ 1.4   \\
CEM-PHX       & 51468   & 250.3 $\pm$ 1.4 & 250.3 $\pm$ 1.4  \\
   \hline            			        
CMUP-CMUP     & 14096   & 251.2 $\pm$ 2.4 & 251.2 $\pm$ 2.4  \\
CMUP-CMU      &  5914   & 261.4 $\pm$ 3.7 & 243.8 $\pm$ 4.5  \\
CMUP-CMP      &  6944   & 244.3 $\pm$ 3.2 & 250.1 $\pm$ 4.2  \\
CMUP-CMX      & 20007   & 247.1 $\pm$ 2.0 & 247.1 $\pm$ 2.0  \\
CMUP-CMIO     &  9099   & 229.2 $\pm$ 2.7 & 240.3 $\pm$ 3.6  \\
CMX-CMX       &  5563   & 250.8 $\pm$ 3.6 & 250.8 $\pm$ 3.6  \\
CMX-CMU       &  3969   & 275.5 $\pm$ 4.7 & 257.0 $\pm$ 5.3  \\
CMX-CMP       &  4280   & 241.8 $\pm$ 3.9 & 247.6 $\pm$ 4.8  \\
CMX-CMIO     &  5771   & 251.0 $\pm$ 3.6  & 263.2 $\pm$ 4.4  \\
 \hline\hline    	  						   
   \end{tabular}
 \end{center}
\label{tab:Zxsec}
\end{table*}

%

\section{\label{sec:Bak}Backgrounds}
We consider four different sources of standard model processes that can
mimic the signature of dilepton plus $\met$ plus 2 or more jets signature:
diboson events ($WW$, $WZ$, $ZZ$ or $W\gamma$), 
Drell-Yan production of tau leptons 
(DY$\rightarrow\tau\tau$), Drell-Yan production of electrons or muons
with additional $\met$ (if the event is an actual Drell-Yan event,
there is no $\met$ so we refer to this as fake $\met$) (DY$\rightarrow ee/\mu\mu$) and QCD production 
of $W$ boson with multiple jets in which one jet is misidentified as a
lepton (``$W$+jet fakes''). 
The two dominant sources of background are
DY$\rightarrow$ $ee$/$\mu\mu$ and $W$+jet fakes.
These two processes have production
cross sections much larger than the $\ttbar$, but they can only contaminate
the $\ttbar$ dilepton signature of two leptons plus jets and large $\met$
when misreconstructions of the event create either some
large fake $\met$ or a jet misidentified as a lepton.
Because it is difficult to use the Monte Carlo simulation to predict the
effect of event misreconstruction in our detector, we estimate the
background from these two processes using data-based methods, as discussed
in Sec.~\ref{sec:fakes} and \ref{sec:DY}, respectively.
The diboson and DY$\rightarrow\tau\tau$ backgrounds
are calculated using Monte Carlo simulation expectation
as described in Secs.~\ref{sec:dib} and \ref{sec:Ztt}. 
Corrections are applied for trigger and lepton 
ID efficiencies following the same procedure described
in Sec.~\ref{sec:Den}.
 
Our strategy for validating 
the background estimation is to compare data
and background estimates in the 0-jet and 1-jet bins, as discussed
in Sec.~\ref{sec:control-samples}.

\subsection{\label{sec:fakes}$W$+jet fakes}

Events with a single $W$ boson plus jets can simulate the dilepton signature
when one of the jets is misidentified as a lepton. 
The $W$+jet fake contamination is calculated in two steps: first 
we extract the probability of a generic QCD jets faking the signatures of
different lepton categories; then we apply these probabilities 
to weight events in the data
containing one and only one high $p_T$ lepton plus jets.

The fake probabilities are 
measured in generic jets from QCD decays by selecting ``fakeable'' leptons, 
which are jets passing minimal lepton identification criteria described below.
We do not consider separately the heavy flavor contribution to our backgrounds because the probability
for a {\it b} or {\it c} quark to become a well reconstructed high $\pt$ lepton is very small.
We define different categories of fakeable lepton,
one per high $p_T$ lepton category in the DIL dilepton selection.

Jets with a large fraction of neutral to charged pion 
production can create signatures with low track multiplicity and 
large energy deposition in the electromagnetic calorimeter, 
thus faking the presence of electrons. 
We define fakeable electrons as tracks of $p_T >$ 20 GeV/$c$
pointing to an electron-like cluster with
energy deposition in the electromagnetic section of the calorimeter 
far exceeding the energy measured in the hadronic  
section, namely with E$_{\rm HAD}/$E$_{\rm EM}<$ 0.125. 
Fakeable electrons are further divided into objects that can fake
CEM or PHX electrons depending on whether their clusters belong to the central
or plug section of the calorimeter. We label them TCEM and TPHX, respectively. 
Fakeable for the non isolated electrons
do not require isolation for the central cluster and are called NCEM.

Jets whose full hadronic activity 
is limited to single charged pions or kaons with a late shower development
or decay in flight might deposit little energy in the calorimeter 
but generate hits in the
muon chambers, thus faking the signature of a muon. 
We define fakeable muons as
good quality tracks of $p_T$ $>$ 20 GeV/$c$ with E/p $<$ 1.  
Depending on which muon sub-detectors these tracks point to, we label as
TCMUP, TCMX, LMIO and LMUO fakeable muons which can fake tight CMUP, 
tight CMX, loose CMIO or loose CMU/CMP muons, respectively. 
Fakeable muons that fail the isolation requirement are put 
together into a single
NMUO category as long as they point to any muon sub-detector.

We select fakeable leptons among generic QCD jets collected in four different
control samples, whose main trigger requirement is the presence of at least
one jet of $E_T^{ Trg}$ $>$ 20, 50, 70 and 100 GeV, respectively. 
The simplification of the jet 
algorithm used in these trigger selection tends to underestimate the energy of 
the offline reconstructed jets. 
To ensure a trigger efficiency of 90\% or greater we require
the trigger jet to have reconstructed $E_T$ greater than
35, 55, 75 and 105 GeV, respectively, in the four jet samples.
The resulting probabilities are labeled 
Jet20, Jet50, Jet70 and Jet100 fake lepton probabilities. 
To minimize real lepton contamination, we require 
that fakeables in the denominator of the fake probability
fail one or more of the standard 
lepton identification cuts. For the numerator instead we
require that the fakeable leptons pass all of the lepton identification 
requirements.
We estimate the contamination of real leptons from
$W$'s, Drell-Yan or dibosons,
using Monte Carlo simulation predictions for the number of events with one lepton 
and at least one jet above the  $E_T$ threshold.

We use the fake lepton probability measured in the Jet50 sample 
as our primary estimator to apply to data events because 
the jet energy spectrum in the Jet50 sample 
is the closest to the energy spectrum of jets in the dilepton 
plus missing E$_{\rm T}$ sample. 
The fake probabilities for different lepton categories show a dependence
on the transverse energy of the fakeable lepton. 
To properly account for difference in 
the $p_T$ spectrum of fakeable leptons in QCD jets vs $W$+jets, we 
calculate fake probabilities in six $p_T$ ranges
as shown in Table~\ref{tab:fake_rate}. 

\begin{figure*}[htbp]
 \begin{center}
  \includegraphics[width=0.80\textwidth, clip]{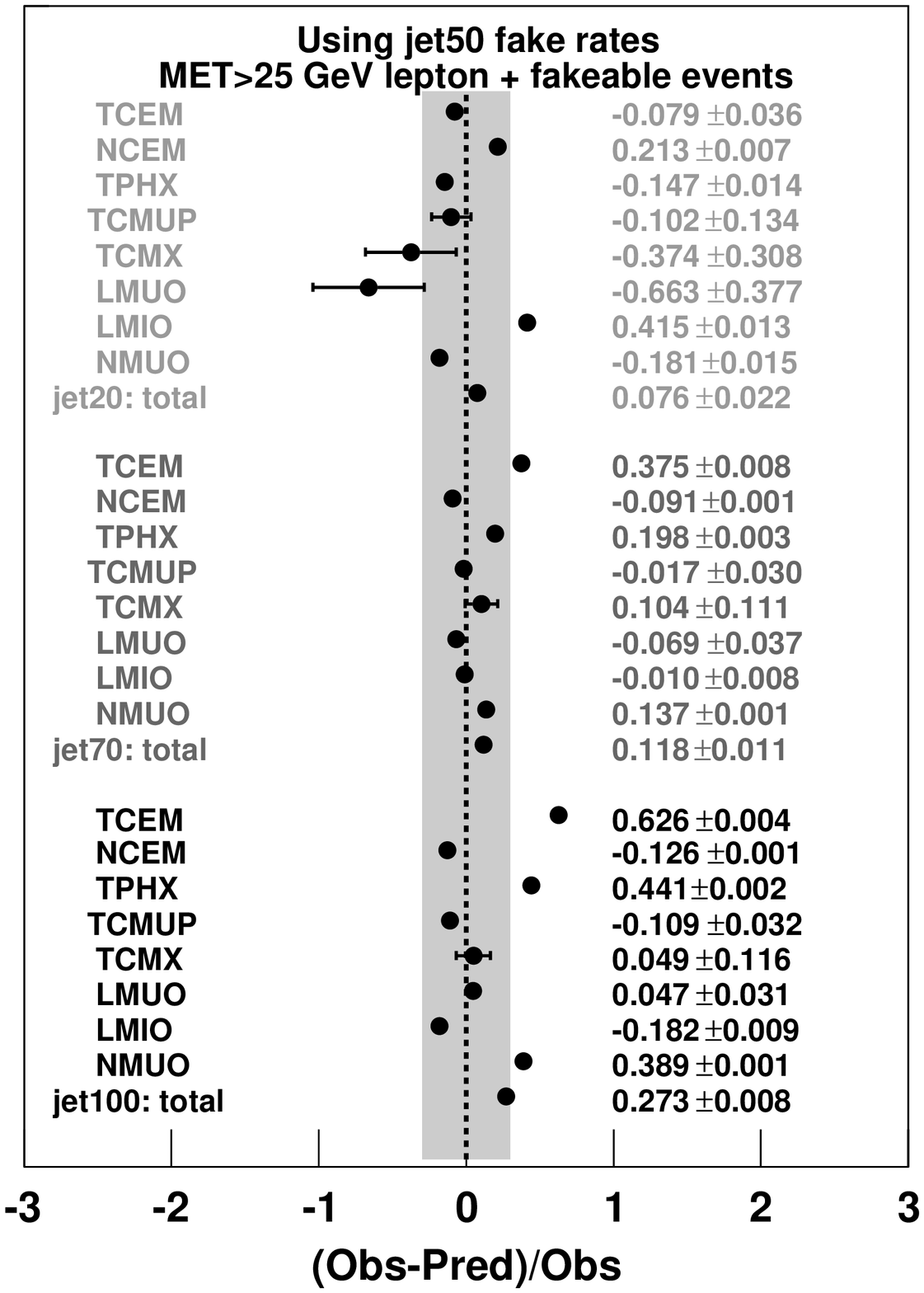}
  \caption{ Ratio of observed total number of fake leptons for each
            fakeable category vs the Jet50-based prediction normalized by the
            number of observed.
            The predictability of the jet50 P$_T$ dependent
            fake rate is good at the 30\% level, as shown
            by the band in the plot. When error bars are not shown they
            are smaller than the dot size.}            
  \label{fig:obs_vs_pred}
 \end{center}
\end{figure*}

The uncertainties on the fake probabilities in Table~\ref{tab:fake_rate} are only 
statistical. Variations in fake probabilities between the
different QCD jet samples are used to estimate a systematic uncertainty in the
lepton fake estimate. Figure~\ref{fig:obs_vs_pred} shows a
comparison between the number of fake lepton events observed 
in the Jet20, Jet70 and Jet100 data sample, after integration over the 
full p$_T$ spectrum, and the number predicted by the Jet50 fake 
probabilities of Table~\ref{tab:fake_rate}.
We assess a 30\% systematic uncertainty on the
ability of the Jet50 fake probabilities to predict electron and muon fake 
contamination in samples with a wide range of jet energy.

\begin{table*}
 \caption{Jet50 fake probabilities vs fakeable lepton $p_T$ for different 
          fakeable categories. The uncertainties are statistical only.
          Due to the definition of fake probability, the denominator can
          fluctuate to be smaller than the numerator in low statistics 
          high $p_T$ bins, hence fake rate values exceeding 100\%.} 
\footnotesize
 \begin{center}
   \begin{tabular}{l c c c c c c}
 \hline\hline
      \multicolumn{7}{c}{Jet50 Fake Probabilities (\%) in $p_T$ range (GeV/$c$) } \\
 \hline
 Fakeable &  [20--30]   &  [30--40]   &  [40--60]   
           &  [60--100]   &  [100--200] &  $\ge 200$  \\
 \hline
 TCEM   & 4.97 $\pm$ 0.09 & 3.68 $\pm$ 0.08 & 2.39 $\pm$ 0.01 
        & 2.88 $\pm$ 0.02 & 3.26 $\pm$ 0.11 & 4.98 $\pm$ 3.44 \\ 
 NCEM   & 0.74 $\pm$ 0.1 & 0.53 $\pm$ 0.01 & 0.58 $\pm$ 0.01 
        & 0.49 $\pm$ 0.01 & 0.80 $\pm$ 0.07 & --- \\
 TPHX   & 12.6 $\pm$ 0.1 & 13.9 $\pm$ 0.1 & 12.7 $\pm$ 0.1 
        & 20.4 $\pm$ 0.2 & 26.8 $\pm$ 0.2 & 65.0 $\pm$ 57.3 \\
 TCMUP  & 0.99 $\pm$ 0.04 & 2.30 $\pm$ 0.10 & 3.82 $\pm$ 0.14 
        & 6.30 $\pm$ 0.28 & 4.13 $\pm$ 0.47 & --- \\
 TCMX   & 0.91 $\pm$ 0.06 & 2.20 $\pm$ 0.15 & 4.74 $\pm$ 0.21                           & 5.09 $\pm$ 0.55 & 0.76 $\pm$ 1.98 & 0.44 $\pm$ 0.25 \\
 LMUO   & 2.48 $\pm$ 0.05 & 2.88 $\pm$ 0.14 & 4.41 $\pm$ 0.23 
        & 5.45 $\pm$ 0.47 & 7.51 $\pm$ 0.64 & 0.41 $\pm$ 0.26 \\
 LMIO   & 21.0 $\pm$ 0.1 & 25.4 $\pm$ 0.3 & 27.8 $\pm$ 0.4 
        & 40.1 $\pm$ 0.1 & 37.0 $\pm$ 1.5 & 107.1 $\pm$ 14.8 \\
 NMUO   & 0.51 $\pm$ 0.01 & 0.36 $\pm$ 0.01 & 0.23 $\pm$ 0.01 
        & 0.25 $\pm$ 0.01 & 0.29 $\pm$ 0.04 & 0.17 $\pm$ 0.18 \\

 \hline\hline
   \end{tabular}
 \end{center}
 \label{tab:fake_rate}
\end{table*}

We define ``lepton+fakeable'' 
as those events in the central high $p_T$ lepton data sets
with one and only one good high $p_T$ lepton, $\metc > $25 GeV and 
a second fakeable object failing at least one standard lepton 
identification cuts.
The fakeable object, which can be from any of the fakeable categories 
defined above, is paired to the good lepton and treated as the second
lepton in the event when calculating any of the kinematic variables
used in the top quark DIL selection, such as
dilepton invariant mass, corrected  $\metc$, and $\Ht$. 
Jets found in a cone of 
$\Delta {\rm R} < 0.4$ around the fakeable lepton are not included in the jet
multiplicity count of that event because those jets are associated with 
the fake lepton in this $W$+jet fake estimation scheme.
The fake lepton contamination
is calculated by weighting each ``lepton+fakeable'' event found in data by the 
fake probability in Table~\ref{tab:fake_rate}.
If more than one fakeable object is found in the event, we pair each
of them to the good lepton and add their single fake contributions.
The fake dilepton background thus calculated contains a statistical
component, which is the
sum of the fake probability uncertainty itself and the statistics of the
``lepton+fakeable'' sample.

As a check,
we compare the same sign ``lepton+fakeable'' prediction to the number of $W$+jet
fakes with same sign dilepton candidates in the signal regions.
We define as fake lepton charge
the charge of the track associated to the fakeable lepton. 
Same sign dilepton candidates are corrected for the presence 
of same sign pairs coming from $t\overline{t}$, 
DY or diboson events that are simulated in our Monte Carlo simulations.
The results of this check are shown in Table~\ref{tab:ss}. 
Although the $\mu\mu$ channel shows deviation at the 3 standard deviation level
for some jet multiplicity bins, the agreement in the final predictions 
over all dilepton categories is at the one standard deviation level.

\begin{table*}
 \caption{Comparison between the same sign dilepton fake background prediction
          using the fake rate tables and the numbers of
          same sign dilepton candidates found in the signal region,
          after MC subtraction of standard model contamination sources.}

\small
 \begin{center}
   \begin{tabular}{lcccc}
 \hline\hline
  Number of SS Dilepton Events      & $ee$                   &  $\mu\mu$       &    $e\mu$        &  all \\
 \hline
Corrected Candidates in 0 jet            & 13.0 $\pm$ 4.4    & 11.3 $\pm$ 3.7   & 14.8 $\pm$ 5.0  & 39.0 $\pm$ 7.6 \\
Predicted Candidates in 0 jet            & 8.1 $\pm$ 2.6     & 7.8 $\pm$ 2.7   & 18.0 $\pm$ 5.6  & 33.9 $\pm$ 10.4\\
 \hline
Corrected Candidates in 1 jet            & 7.8 $\pm$ 3.5      & 0.5 $\pm$ 1.4    & 24.9 $\pm$ 5.6  & 33.2 $\pm$ 6.7  \\
Predicted Candidates in 1 jet           & 5.1 $\pm$ 1.6      & 6.9 $\pm$ 2.2    & 25.4 $\pm$ 7.8  & 37.3 $\pm$ 11.4  \\
 \hline
Corrected Candidates in 2 or more jets    & 5.0 $\pm$ 2.7    & 0.7 $\pm$ 1.0     & 24.7 $\pm$ 5.2  & 30.4 $\pm$ 5.9 \\
Predicted Candidates in 2 or more jets   & 3.8 $\pm$ 1.2    & 7.5 $\pm$ 2.4     & 24.5 $\pm$ 7.6  & 35.9 $\pm$ 11.0  \\
 \hline\hline

   \end{tabular}
 \end{center}
 \label{tab:ss}
\end{table*}

\subsection{\label{sec:DY} Drell-Yan to $ee$/$\mu\mu$ Background}

The contamination from $Z/\gamma^* \rightarrow ee/\mu\mu$ decays
is calculated using a combination of data based and MC based predictions.
We define DIL data samples enriched in DY events after the L-cut
by inverting the $Z$-veto cut and extrapolating
the remaining DY contamination in the signal region by using
the relative contribution of 
$Z/\gamma^* \rightarrow ee/\mu\mu$ decays passing and failing
the $Z$-veto cut as predicted from MC.
The $Z$-veto cut (see Sec.~\ref{sec:Data}) requires that the dilepton
invariant mass be outside the $Z$ window region of 76 to 106 GeV/$c^2$, 
or, if inside, that the event have missing $E_T$ 
significance $\metc /$$\sqrt{{ E}^{ sum}_{ T}} >$ 4~{GeV}$^{(1/2)}$.

\par
We calculate the DY$\rightarrow $ee$/\mu\mu$ contamination 
as the sum of two contributions, one outside the $Z$ window region, N$_{\rm out}$, 
and one inside the $Z$ window with high MetSig, N$_{\rm high}$.
The first contribution is calculated as:
\begin{equation}
   {\rm N_{\rm out}} =  {\rm R_{\rm out/in}} 
                {\rm ( N_{\rm in}^{\rm DT} - N_{\rm in}^{\rm BKG} )} \\ 
  \label{eq:DY_bkgr_1}
\end{equation}\noindent
where N$_{\rm in}^{\rm DT}$ and N$_{\rm in}^{\rm BKG}$ represent
the number of events inside the $Z$-window
passing the L-cut in data
and in non-DY MC background predictions, respectively.
R$_{\rm out/in}$ is the ratio of $Z$$/\gamma^* \rightarrow ee/\mu\mu$ events 
outside to inside the $Z$ window predicted by the {\sc alpgen}~\cite{alpgen}
Monte Carlo generator.

The second contributions is calculated as:
\begin{equation}
   {\rm N_{\rm high}} = {\rm R_{\rm high/low}}
                {\rm ( N_{\rm low}^{\rm DT} - N_{\rm low}^{\rm BKG} )} 
  \label{eq:DY_bkgr_2}
\end{equation}
\noindent
where  N$_{\rm low}^{\rm DT}$ and N$_{\rm low}^{\rm BKG}$ represent
the events inside the $Z$-window passing the L-cut
with MetSig $<$ 4~{GeV}$^{(1/2)}$
for data and for non-DY MC background predictions, respectively.  
R$_{\rm high/low}$ is the ratio of events passing/failing
the MetSig $>$ 4~{GeV}$^{(1/2)}$ cut predicted by {\sc alpgen}.
Table~\ref{tab:DY_bkgr} summarizes the inputs to Eqs.~(\ref{eq:DY_bkgr_1})
and (\ref{eq:DY_bkgr_2})
and the final values of N$_{\rm out}$ and 
N$_{\rm high}$ for each jet multiplicity bin.
For the calculation of $\ttbar$ contribution to ${\rm N^{\rm BKG}}$ we use the
prediction of 6.7~pb for the cross section. 
We later correct this iteratively to the value measured in the data.

\begin{table*}
\caption{ Inputs to Eqs.~(\ref{eq:DY_bkgr_1}) and (\ref{eq:DY_bkgr_2}) and
          for each dilepton flavor and jet multiplicity. 
          N$_{\rm out}$ and N$_{\rm high}$ are the final values of
          the DY$\rightarrow ee$ and $\mu\mu$ background contamination
          outside the $Z$ peak region and inside the $Z$ peak region with
          high MetSig, respectively.}
\footnotesize
  \begin{center} 
   \begin{tabular}{lcccccc}
   \hline\hline
          & \multicolumn{2}{c}{ 0-jet} 
	  & \multicolumn{2}{c}{ 1-jet} 
          & \multicolumn{2}{c}{$\ge$2-jets}  \\
      &  $ee$  & $\mu\mu$   &  $ee$  & $\mu\mu$  &  $ee$  & $\mu\mu$  \\
                         &      &     &     &     &   &      \\
   \hline
${\rm N_{\rm in}^{\rm DT}}$ &  78  &  45    &  76  &  58   &  73 &  43 \\  
${\rm N_{\rm in}^{\rm BKG}}$ & 29.6$\pm$1.4  & 20.1$\pm$1.0   & 
                               16.5$\pm$1.6  & 11.9$\pm$1.4   &  
                               17.0$\pm$1.2  & 16.8$\pm$1.0   \\

${\rm R_{\rm out/in}}$    &  0.39$\pm$0.05  & 0.45$\pm$0.09   &   
                             0.31$\pm$0.05  & 0.31$\pm$0.06   &
                             0.32$\pm$0.02  & 0.26$\pm$0.02   \\
   \hline
${\rm N_{\rm out}}$     &  19.1$\pm$3.7  &   11.3$\pm$3.2   & 
                           18.2$\pm$2.8  &   14.2$\pm$2.5   &  
                           18.2$\pm$2.8  &    6.8$\pm$1.7   \\
                         &      &     &     &     &   &     \\
   \hline
${\rm N_{\rm low}^{\rm DT}}$  &  65 &  37     &  69  &  54    &  65 & 39     \\
${\rm N_{\rm low}^{\rm BKG}}$  &  12.6$\pm$0.7  &  9.5$\pm$0.6    &   
                                   5.9$\pm$0.5  &  5.5$\pm$0.5    &  
                                   6.7$\pm$0.2  &  8.0$\pm$0.5  \\

${\rm R_{\rm high/low}}$       &  0.026$\pm$0.009  & 0.010$\pm$0.005  & 
                                  0.049$\pm$0.006  & 0.049$\pm$0.011  & 
                                  0.040$\pm$0.003  & 0.040$\pm$0.004  \\
   \hline
${\rm N_{\rm high}}$           &   1.23$\pm$0.21    & 0.26$\pm$0.06   & 
                                   2.85$\pm$0.41    & 2.34$\pm$0.36   & 
                                   2.16$\pm$0.32    & 1.22$\pm$0.25   \\
   \hline\hline
   \end{tabular}
 \end{center}
 \label{tab:DY_bkgr}
\end{table*}

\par
The DY contamination in the signal sample
is extracted from the N$_{\rm out}$ and N$_{\rm high}$ estimates 
in the $\ge 2$ jet bin, corrected for 
the efficiency of the $\Ht > 200$~GeV and of the
opposite sign lepton cuts.
The combined efficiency for these two cuts
is calculated using {\sc alpgen} simulated $Z$ samples
and shown as $\epsilon_{\rm OS}$ in Table~\ref{tab:HT_DY_bkgr}.

\par
The contamination of $Z/\gamma^*\rightarrow \mu\mu$ 
to e$\mu$ events comes mostly from cases
where one of the final state muon radiates a very energetic
photon. These photons, which are almost collinear to the
muon, deposit their energy in the EM calorimeter and
produce a cluster which is
associated to the original muon track and fakes the electron signature.
The missed muon gives rise to a sizable $\met$ in the event,
curtailing the effectiveness of the L-cut and $Z$-veto to reject them.
As no data based control sample is available for this contamination, we
estimate it using Monte Carlo simulation predictions.

\begin{table*}
\caption{ $\Ht$ and opposite sign cut efficiency for the 
           DY$\rightarrow ee$ and $\mu\mu$ background contamination	
           in $\ge 2$ jet region. The efficiency is calculated separately
           for events outside the $Z$ peak region passing the
           L-cut, and for events inside the $Z$ peak region also 
           passing the MetSig $>$ 4~{GeV}$^{(1/2)}$ cut.} 
\begin{center} 
   \begin{tabular}{c c c}
   \hline\hline
 $\epsilon_{\rm H_{T},OS}$              &   $ee$        &  $\mu\mu$     \\     
   \hline
for N$_{\rm out}$ events     & 0.54$\pm$0.03 & 0.60$\pm$0.05 \\
for N$_{\rm in}$~ events     & 0.95$\pm$0.01 & 0.99$\pm$0.01 \\ 
   \hline\hline
   \end{tabular}
 \end{center}
 \label{tab:HT_DY_bkgr}
 \end{table*}

\subsection{\label{sec:dib}Diboson Background}

The diboson processes, $WW$, $WZ$, $ZZ$ and $W\gamma$, 
can mimic the signature of the $\ttbar$ signal via
different mechanisms, with real leptons and $\met$ from $W$ and $Z$ decays
and jets produced by boson hadronic decays or initial and final state radiation.
For $WW$ events,
the two leptons and the $\met$ are produced when both $W$'s decay 
semi-leptonically
but the jets require some hadronic radiation external to the diboson system.
For $WZ$ and $ZZ$ events, the two leptons come from 
the $Z$ boson while the other $W$ or $Z$ boson provides the jets via their  
hadronic decays. As these decays do not contain any neutrino, some mechanism 
to produce fake $\met$ is required.
Finally for $W\gamma$ events, one lepton plus $\met$ is generated from the
semi-leptonc $W$ decay while the second lepton is 
produced from an asymmetric $\gamma$ conversion
in which one of the two electrons has little energy and is caught spiralling
inside the central drift chamber. Like in the 
$WW$ case, the $W\gamma$ system is accompanied by hadronic jets. 
Events involving $W$+jets fake leptons, with a real lepton
from $W$ boson paired to a fake lepton from the hadronic decays of
the other boson, are removed from the MC to avoid double
counting.

Only $WW$ background contribute to the $ee$, $\mu\mu$ and $e\mu$ final states
in the same proportion as the $\ttbar$ signal.
Diboson processes involving a $Z$
contribute preferentially to the same flavor lepton categories. 
$W\gamma$ events do not contribute any
background to the $\mu^+\mu^-$ category given the negligible probability that 
the photon will convert to a muon pair.

The $WW$, $WZ$ and $ZZ$ processes are simulated with the {\sc pythia} Monte Carlo
generator. Their production cross
section is taken from the latest next-to-leading order (NLO) 
MCFM version~\cite{NLO} and
CTEQ6~\cite{CTEQ} PDF predictions to be $\sigma_{WW}$ = 12.4 $\pm$
0.8 pb, $\sigma _{WZ}=3.7 \pm 0.1$ pb. For the $Z$$Z$ events, a
cross section  $\sigma_{ZZ} = 3.8$ pb is assumed with an
uncertainty of 20\%. 
$W\gamma$ decays are simulated with the {\sc BAUR}
Monte Carlo generator~\cite{baur}. The leading order (LO) 
production cross section of
$\sigma_{W \gamma}$ = 32 $\pm$ 3 pb is assumed, and multiplied by a K-factor of 1.36~\cite{baur2} 
to correct for NLO effects.
The $W\gamma$ Monte Carlo generator acceptance prediction is multiplied by
a conversion inefficiency scale factor of 1.15 $\pm$ 0.35 to correct 
for the imperfect simulation of the tracking variables used 
in the conversion identification algorithm.

\par
Monte Carlo generators do not correctly model 
the jet production from hadronic radiation,
as is seen by comparing the jet multiplicity spectra of data 
and MC predictions for $ee$ and $\mu\mu$ events in the $Z$ peak region. 
Data, even after correcting the jet multiplicity spectrum for other
SM contributions, have higher fractions of events in the 2 or more jet
bins compared to predictions.
We calculate jet multiplicity scale factors C$_{\rm Nj}$ 
as ratios of data and MC events in each jet bin, 
after normalizing the MC to the 
number of data in the $Z$ peak region. These scale factors,
shown in Table~\ref{tab:JET_SF}, are used to correct the 
jet multiplicity of $WW$ and $W\gamma$ events. 
A 5\% systematic uncertainty on this correction is assessed by comparing
jet multiplicity scale factors calculated with different generators.

\begin{table}
 \caption{ Jet multiplicity scale factors for
           $Z\to$ $e^+e^-$ and $Z \to\mu^+\mu^-$ events
           in the 0-jet bin (C$_{\rm 0j}$), 1-jet bin (C$_{\rm 1j}$),
           and $\ge$ 2-jet bin (C$_{\rm 2j}$), respectively.
           The last column is the weighted average of the two same flavor $Z$
           samples and it is used as the correction factor for $e\mu$ 
           reconstructed events. The uncertainties shown here
           are statistical only.}
 \begin{center} 
   \begin{tabular}{l c c c}
   \hline\hline
 	&  \multicolumn{3}{c}{Jet multiplicity Scale Factor}  \\
   \hline
        &   $e^+e^-$   &  $\mu^+\mu^-$   &   $\ell^+\ell^-$ \\
   \hline
C$_{\rm 0j}$   & 1.017 $\pm$ 0.010 & 0.999 $\pm$ 0.011 & 1.010 $\pm$ 0.010 \\
C$_{\rm 1j}$   & 0.918 $\pm$ 0.012 & 0.991 $\pm$ 0.012 & 0.948 $\pm$ 0.008 \\
C$_{\rm 2j}$   & 1.056 $\pm$ 0.020 & 1.123 $\pm$ 0.020 & 1.082 $\pm$ 0.014 \\
   \hline\hline
   \end{tabular}
 \end{center}
\label{tab:JET_SF}
\end{table}

\subsection{\label{sec:Ztt} Drell-Yan $ \rightarrow \tau\tau$ Background}

$Z/\gamma ^* \rightarrow \tau^{+}\tau^{-}$ decays
are simulated with the {\sc alpgen} generator. 
These events can fake the dilepton plus $\met$ plus 2 or more jets 
signature when both $\tau$'s decay semi-leptonically to 
$\ell^{+} {\nu}_{\ell} \bar{\nu_{\tau}}\ell^{-}\bar{\nu}_{\ell} \nu_{\tau}$ and jets from initial and final 
state radiation are present. 
The contamination from this process is expected to contribute equally to the 
$e^+e^-$ and $\mu^+\mu^-$ categories and to be twice as big 
in the $e^{\pm}\mu^{\mp}$ channel. 
The neutrinos from the semi-leptonic $\tau$ decays tend to have lower energy
than the neutrinos in the $\ttbar$ dilepton sample 
and align along the direction of the leptonic decay when the $Z$ recoils
against the external jets. Hence a big fraction
of the  $Z/\gamma ^* \rightarrow \tau\tau$ events are removed by the L-cut,
the cut on the event $\metc > 25$~GeV or $\metc > 50$~GeV 
in case any lepton or jet is closer than
20$\rm ^o$ to the $\met$ direction (see Sec.~\ref{sec:Data}).

\par

The final contamination from this process is estimated using a Monte Carlo simulation
and assumes a $Z \rightarrow$ $\tau\tau$ cross section of 251.6$^{+2.8}_{-3.1}$ pb~\cite{Zcross}.
The simulated samples are generated using {\sc alpgen} generator~\cite{alpgen} 
that has built-in matching of the number of jets, coupled with {\sc pythia}~\cite{pythia} 
for the shower evolution and {\sc evtgen}~\cite{evtgen} for the heavy-flavor hadron decays.
All simulated events were run through the full CDF detector simulation.
To correct for NLO effects, this value is further multiplied by a K-factor of 1.4~\cite{DY}.
The MC predictions in the different jet bins are finally rescaled
by the C$_{\rm Nj}$ scale factors, as discussed in Sec. V C.




\section{\label{sec:Sys}Systematic Uncertainties}

The systematic uncertainty for the cross section measurement has 
two main contributions: systematics in the $\ttbar$ dilepton acceptance
and systematics in the background estimation.
We distinguish between
the uncertainties affecting only the signal or the background
from the uncertainties common to both. 

\par
For the signal acceptance, we consider systematic uncertainties
coming from different MC generators,   
different assumed amounts of initial (ISR) and final (FSR) state radiation
in the {\sc pythia} Monte Carlo, calculated by comparing data to expectations 
for the $\pt$ spectrum of the dilepton system in Drell-Yan events and 
for the kinematic distributions of the underlying events
and different parton distribution functions (PDF).
These sources are uncorrelated from each other. The two remaining 
and largest sources of acceptance systematics are common to signal 
and background Monte Carlo simulation predictions. 
They arise from uncertainties in the lepton identification (ID)
scale factors and jet energy scale (JES).
Comparing the lepton ID scale factors calculated for $Z$ events with 
0, 1 and $\ge$2 jets, we derive a systematic associated to the
Monte Carlo generator acceptance correction of 2\%. This is added in quadrature
to the 1\% systematic uncertainty on the acceptance correction procedure
derived from measurement of the $Z$ cross section in different dilepton
channels (see Sec.~\ref{sec:Den}), for a total systematics on the 
lepton ID correction of 2.2\%.
The JES uncertainty 
is calculated by measuring the shift in acceptance due 
to varying the jet energy scale correction applied to each jet 
in the event by $\pm$1 standard deviation of its systematic uncertainty.
Table \ref{tab:sys} summarizes the systematic uncertainties in the 
$\ttbar$ acceptance separated by the contributions which are independent
and contributions which are common with the systematic uncertainty 
in the background prediction.
Common contributions affect both the numerator and the denominator
of Eq.~(\ref{eq:Xsec}) used to calculate the final $\ttbar$ cross section.
Their correlation is taken into account when calculating
the systematic uncertainty on the measured $\ttbar$ cross section.

\begin{table}[htbp]
\caption{Summary of systematic uncertainties affecting the $\ttbar$ 
         acceptance: the entries in the top part of the table are treated as
         uncorrelated and added in quadrature when calculating their
         contribution to $\ttbar$ cross section systematic uncertainty
         via the denominator of Eq.~(\ref{eq:Xsec}); the two 
         entries in the bottom part of the table are also added in
         quadrature but their correlation with the systematic uncertainty
         in the background prediction, which appears in the numerator
         of Eq.~(\ref{eq:Xsec}), is taken into account when 
         calculating the final systematic uncertainty in the $\ttbar$
         cross section.}
 \begin{center}
  \begin{tabular}{l c}
\hline\hline
Source & Systematic Error (\%) \\
\hline
MC Generator & 1.5 \\
ISR & 1.7 \\
FSR & 1.1 \\
PDF & 0.8 \\
\hline
Lepton ID  & 2.2 \\
Jet Corrections & 3.2\\
 \hline\hline
  \end{tabular}
 \end{center}
\label{tab:sys}
\end{table}

\par
Uncorrelated sources of systematic uncertainties affecting the
backgrounds are the 30\% systematic uncertainty on the fakes contamination, 
the 30\% uncertainty on the conversion inefficiency scale factor
affecting the $W\gamma$ contamination and the theoretical 
uncertainties, ranging from 2\% to 10\%, on the production cross sections 
of diboson and $Z$$\to \tau\tau$ processes.
Although large, each of these systematics affects only a fraction of the
total background. Finally,
a systematics common to most Monte Carlo generator predictions of background
processes with jet production from QCD radiation comes from the 5\%
uncertainty in the C$_{\rm Nj}$ correction factors of 
Sec.~\ref{sec:dib}.

\section{\label{sec:control-samples}Control Samples}

\begin{figure*}[htbp]
 \begin{center}
  \begin{tabular}[t]{cc}
    \includegraphics[width=0.45\textwidth, clip]{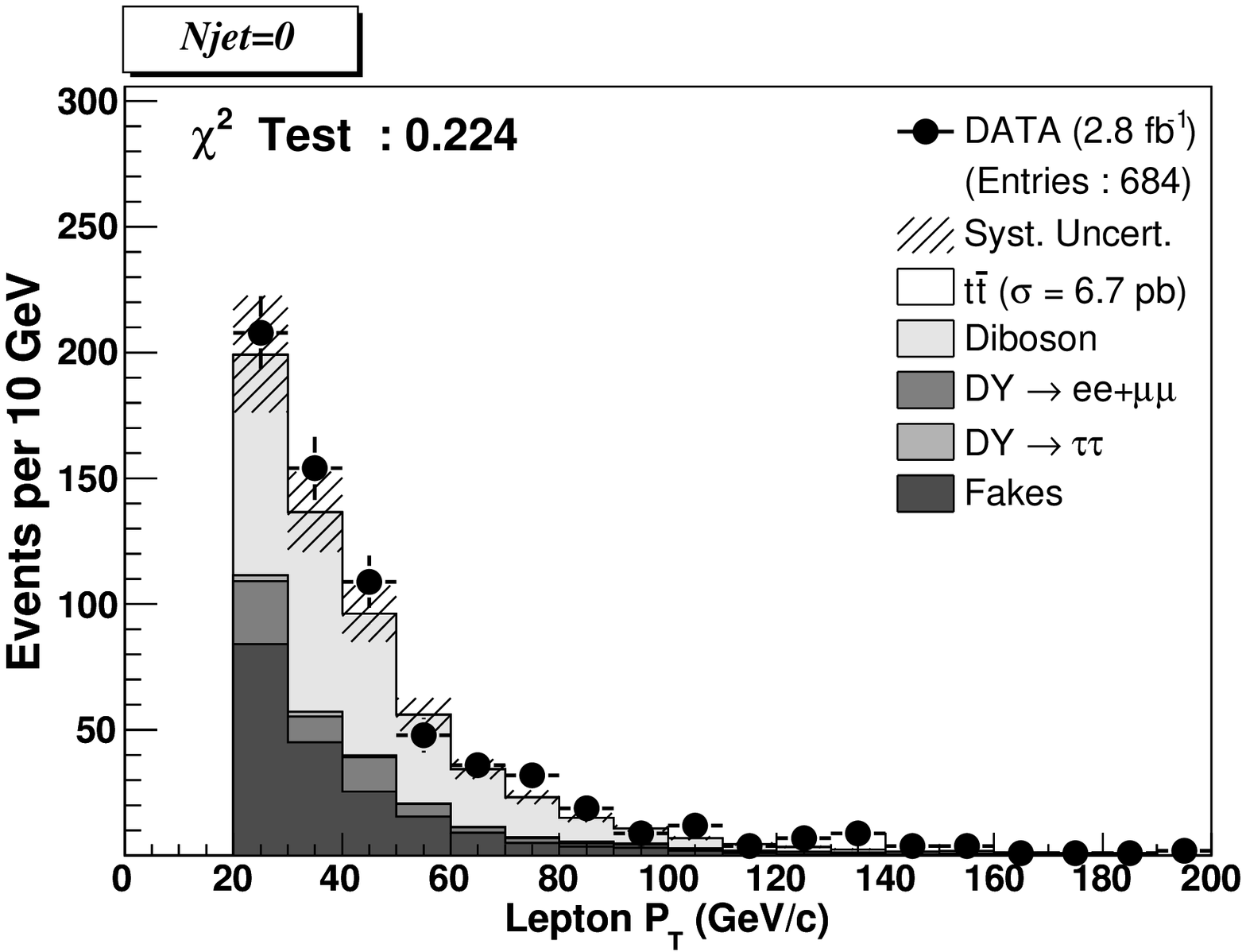} &
    \includegraphics[width=0.45\textwidth, clip]{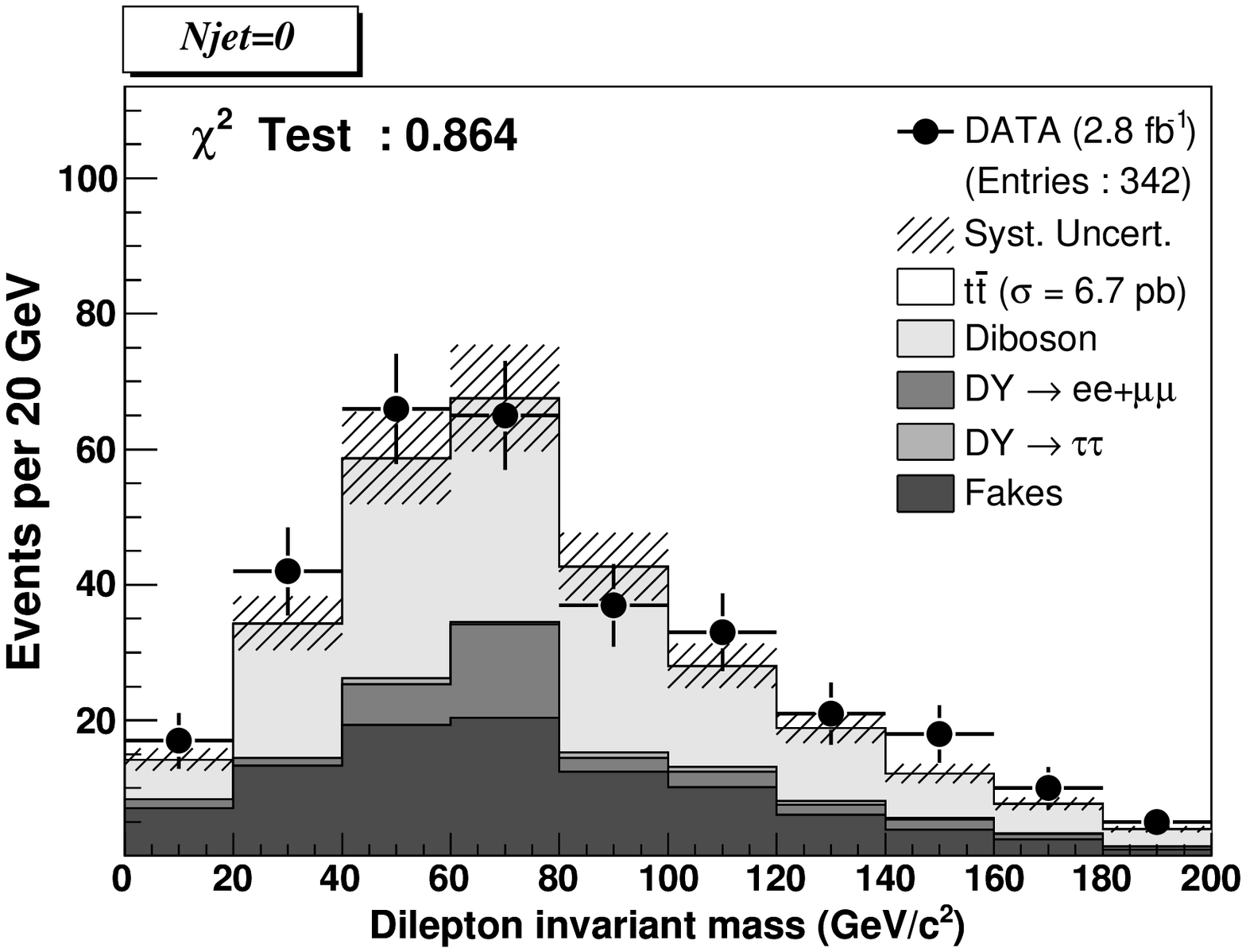} \\
    \includegraphics[width=0.45\textwidth, clip]{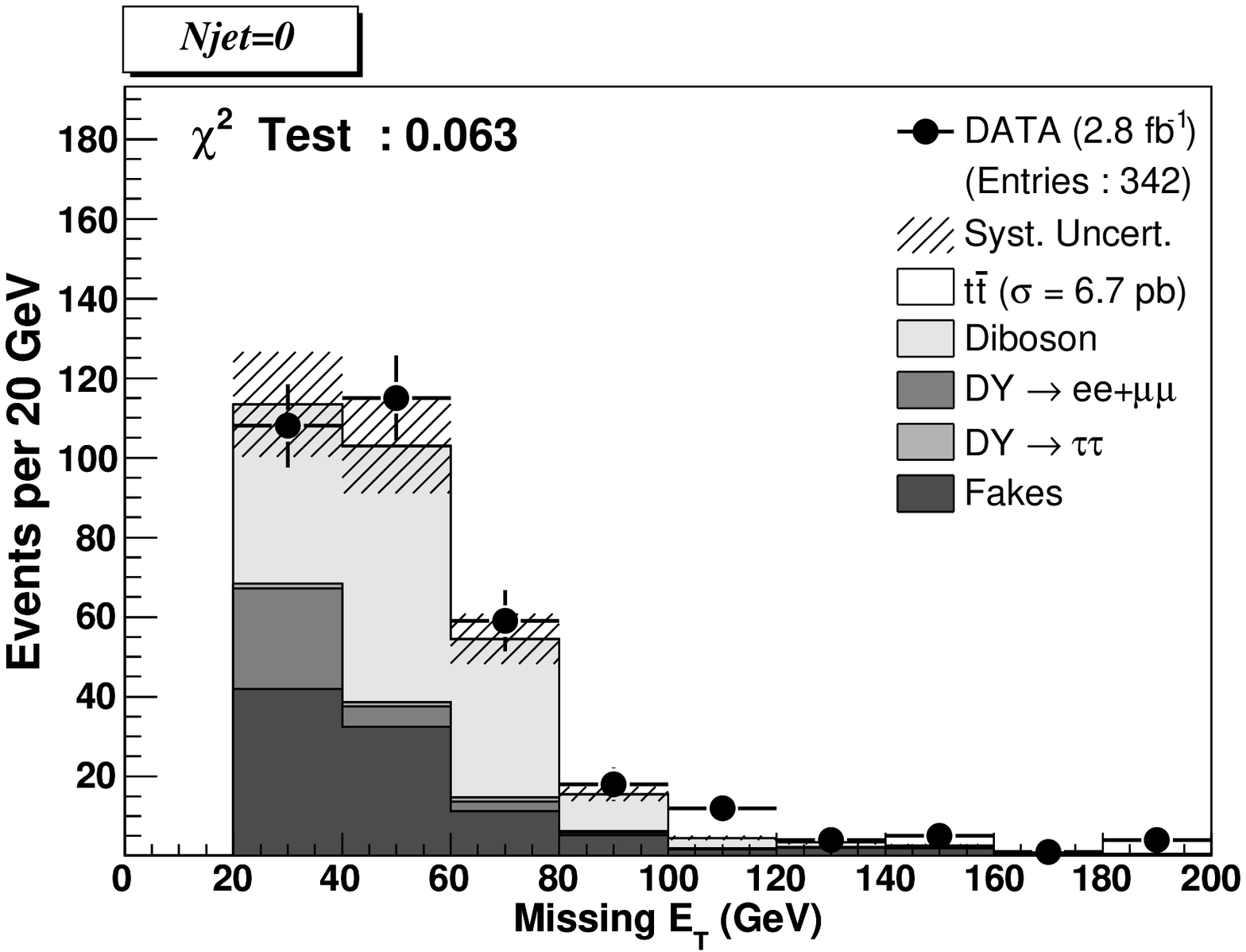} &
    \includegraphics[width=0.45\textwidth, clip]{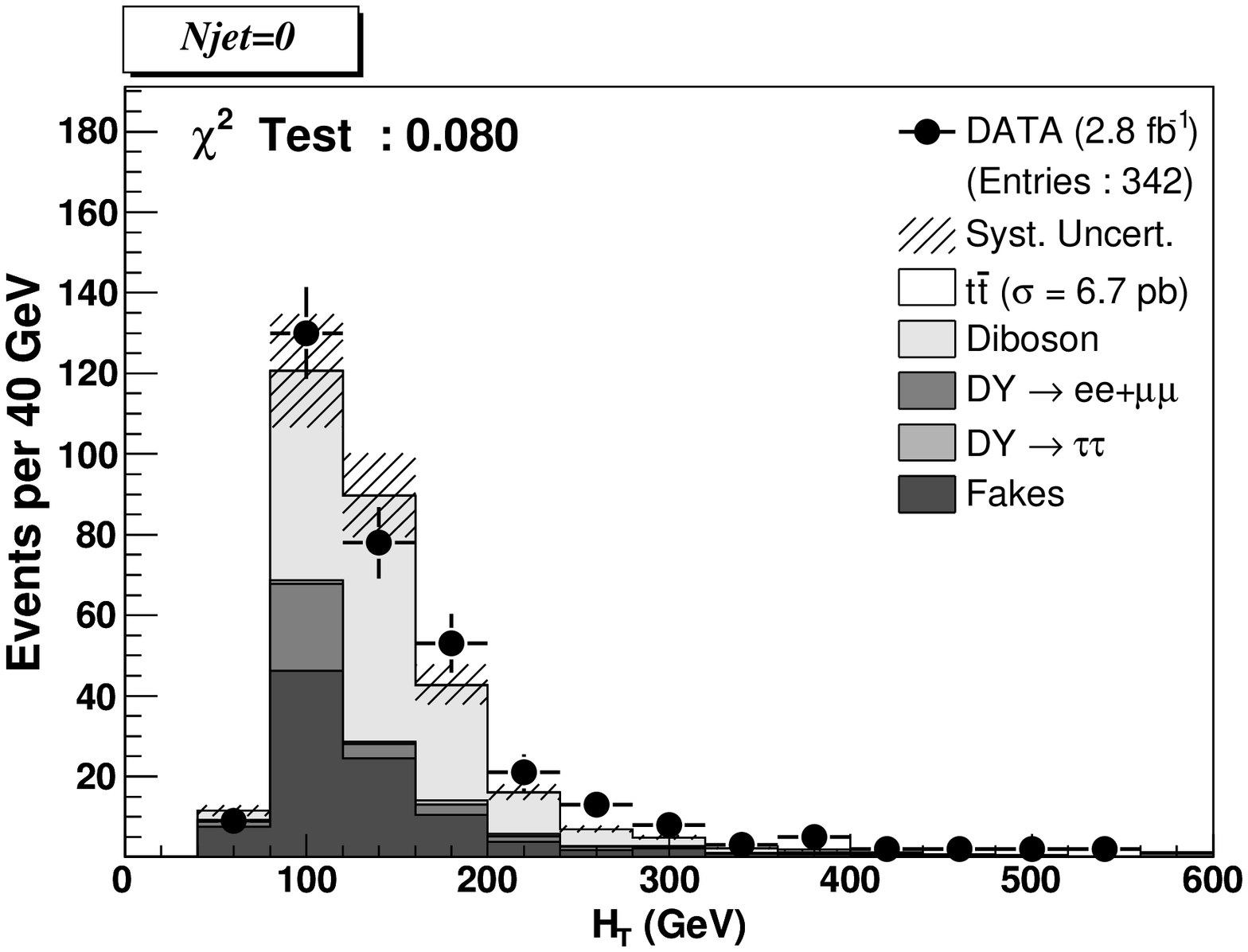} \\         
  \end{tabular}
 \end{center}
  \caption{ Background and top quark signal predictions 
            overlaid on the data for 0-jet events in 2.8~fb$^{-1}$.
            From top left to bottom right: 
            Two leptons transverse energy spectrum,
            the dilepton invariant mass, the event $\met$ and H$_T$. 
            The hatched area represents the uncertainty in the
            total background estimate.} 
  \label{fig:0jet_2}
\end{figure*}

\begin{figure*}[htbp]
 \begin{center}
  \begin{tabular}[t]{cc}
    \includegraphics[width=0.45\textwidth, clip]{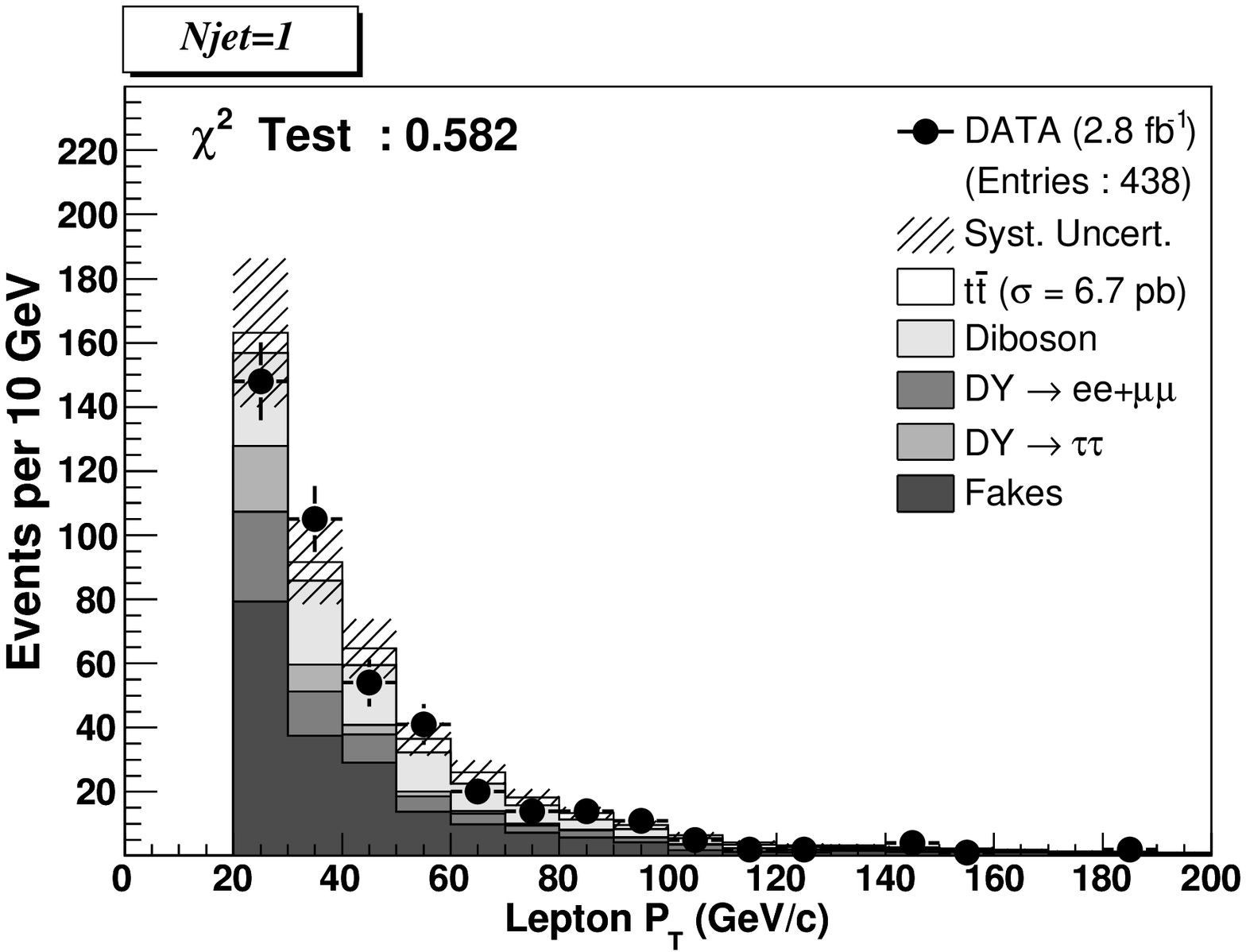} &    
    \includegraphics[width=0.45\textwidth, clip]{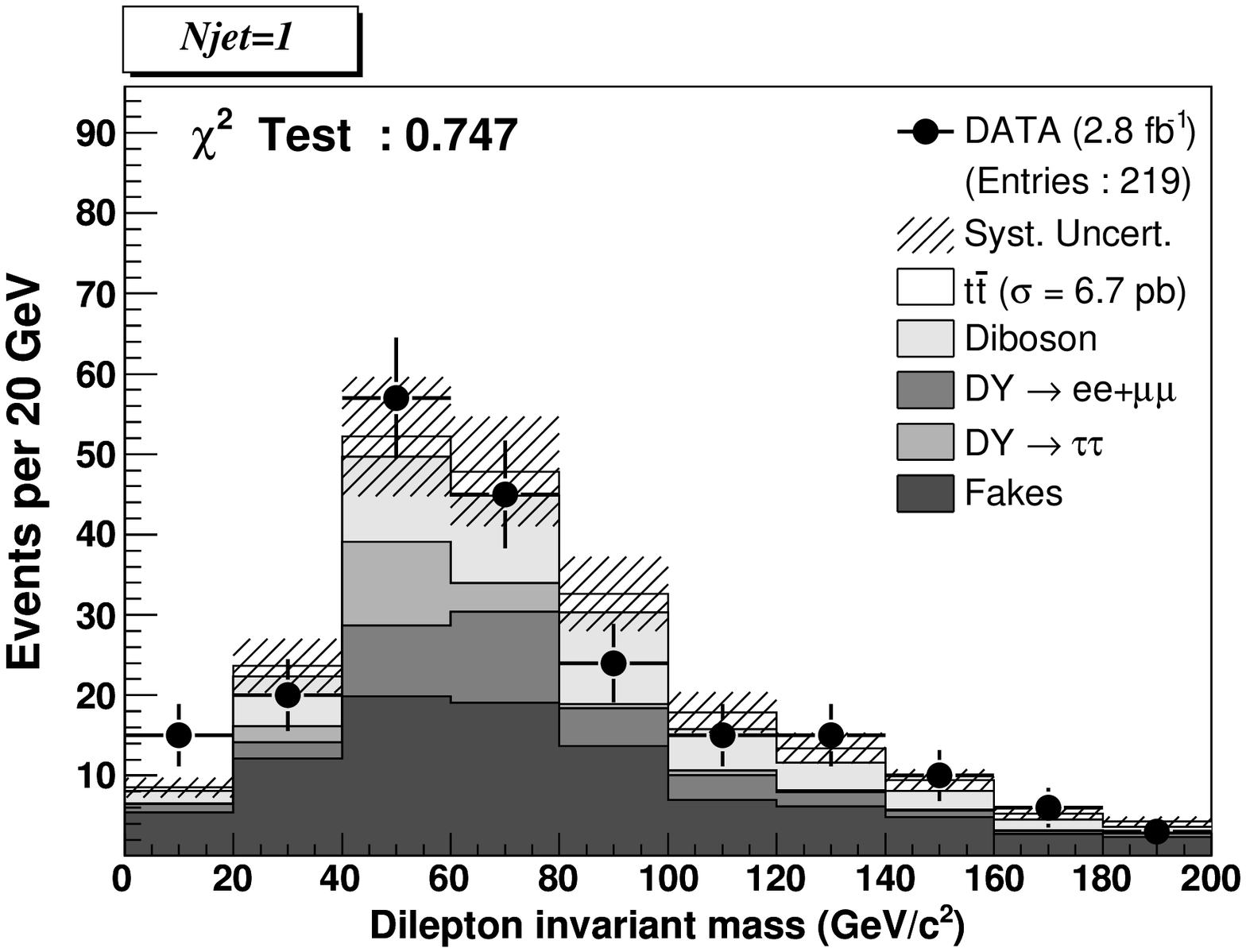} \\  
    \includegraphics[width=0.45\textwidth, clip]{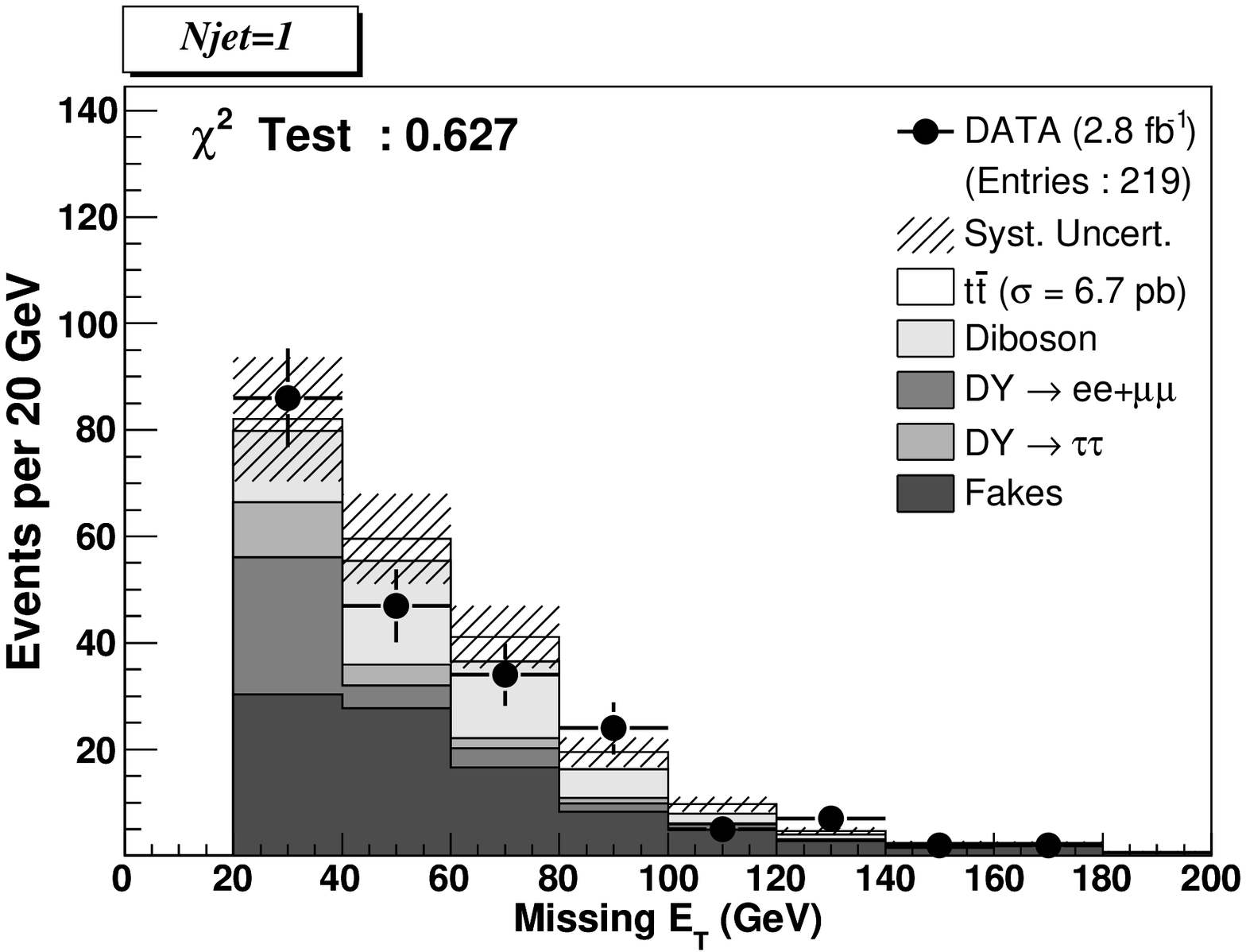} &	    
    \includegraphics[width=0.45\textwidth, clip]{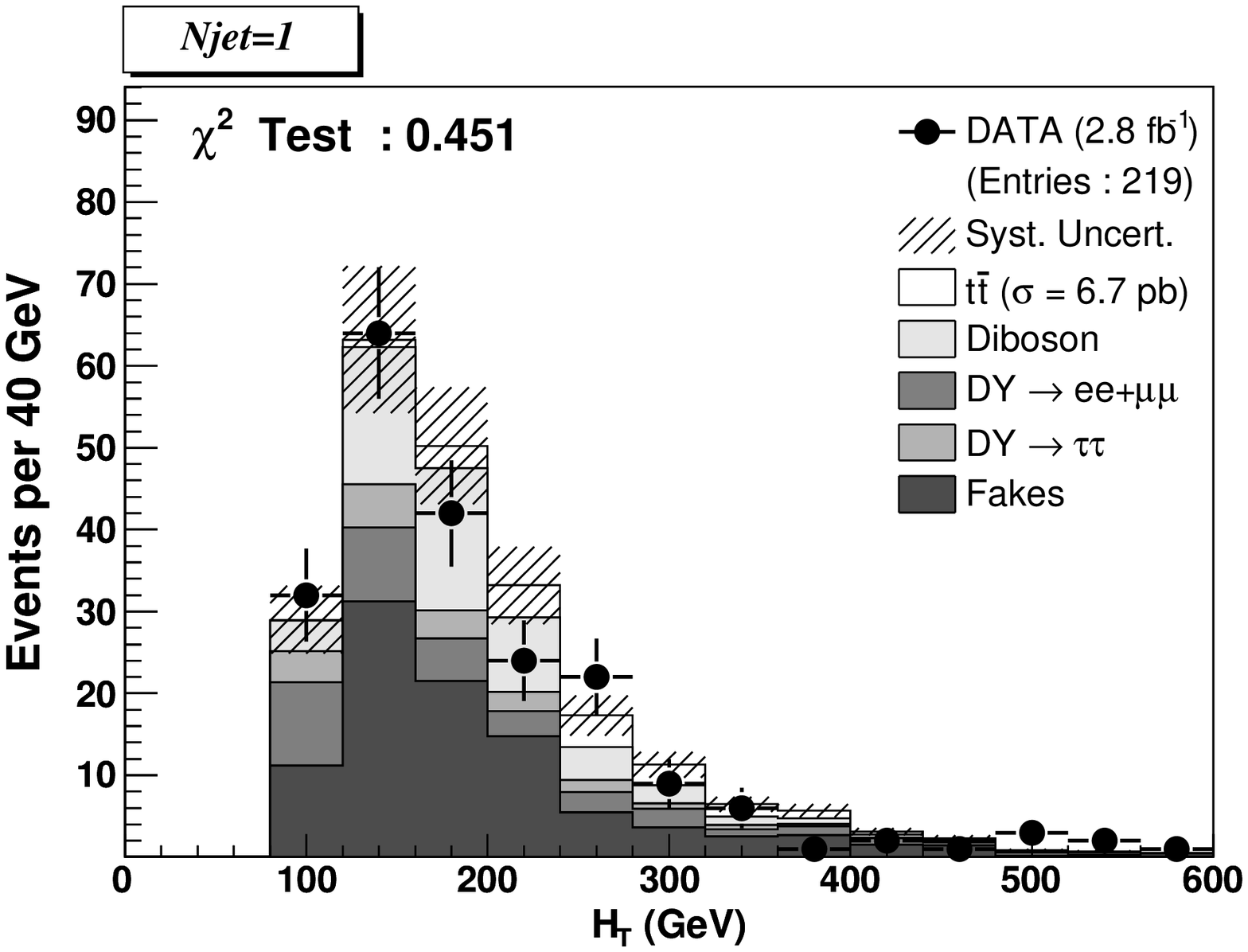} \\         
  \end{tabular}
 \end{center}
  \caption{ Background and top quark signal predictions 
            overlaid on the data for 1-jet events in 2.8~fb$^{-1}$ 
            From top left to bottom right: 
            Two leptons transverse energy spectrum,
            the dilepton invariant mass, the event $\met$ and H$_T$. 
            The hatched area represents the uncertainty in the
            total background estimate.} 
  \label{fig:1jet_2}
\end{figure*}

We use dilepton events passing both the L-cut and $Z$-veto cut, but with
only 0 or 1 jets as control samples for the background calculation.
The $\ttbar$ contamination to the 0 jet samples is negligible and the
contribution to the 1 jet sample is small.
The results for the number of observed and expected events 
in the various dilepton flavor categories
are shown in Table~\ref{tab:cntrl_bkgr_2fb}.
The $\ttbar$ contribution has been calculated assuming
a production cross section of 6.7~pb.
For the 0-jet bin data, the single largest contribution is from 
diboson production, followed by $W$+jet fakes and DY production;
for the 1-jet bin data the single largest contribution is 
from $W$+jet fakes, followed by diboson and DY production.
Figures~\ref{fig:0jet_2} and~\ref{fig:1jet_2}
show the 0-jet and 1-jet data overlaid on top of the background 
and $\ttbar$ predictions
for four different kinematic distributions:
single lepton p$_{T}$, dilepton invariant mass, 
event missing transverse energy $\met$ and 
total scalar transverse energy H$_{T}$.
The largest deviation, still below the 2 standard deviation level, 
is in the $N_{jet}$ = 0 control sample for the $\mu\mu$ channel. 
Overall the data are in good agreement with the expectations of background plus signal ($\ttbar$).
The agreement is quantified 
in terms of the probability for
the $\chi^{2}$/ndf distribution and 
shown as ``$\chi^{2}$ Test'' on the figures.

\begin{table*}
\caption{ Summary table, by lepton flavor content,
          of background estimates, $\ttbar$ predictions 
          and observed events in data corresponding to 
          an integrated luminosity of 2.8~fb$^{-1}$ 
          for the 0 jet (top) and 1 jet (bottom) bins, respectively.
          The quoted uncertainties are the sum of the statistical
          and systematic uncertainty.}
  \begin{center}
  \begin{tabular}{l c c c c}
  \hline\hline
\multicolumn{5}{c}{ N$_{\rm jet}$ = 0 Control Sample per Dilepton Flavor Category}\\
  \hline
 Source          &    $ee$         & $\mu\mu$   &   $e\mu$    &   $\ell\ell$ \\
  \hline

 $WW$  &  36.36$\pm$3.26&  30.10$\pm$2.71&  76.40$\pm$6.76&  142.87$\pm$12.57 \\
 $WZ$  &  2.88$\pm$0.21&  4.56$\pm$0.31&  4.21$\pm$0.29&  11.65$\pm$0.74 \\
 $ZZ$  &  4.13$\pm$3.19&  4.25$\pm$3.28&  0.41$\pm$0.32&  8.79$\pm$6.78 \\
 $W\gamma$ &  14.26$\pm$4.90&  0.00$\pm$0.00&  13.77$\pm$2.63&  28.03$\pm$7.11 \\
 DY$\rightarrow\tau\tau$     &  0.95$\pm$0.26&  0.79$\pm$0.23&  2.15$\pm$0.39&  3.89$\pm$0.58 \\
 DY$\rightarrow ee+\mu\mu$   &  20.32$\pm$3.94&  11.59$\pm$3.34&  8.13$\pm$1.33&  40.04$\pm$5.77 \\
 $W$+jet fakes &  18.72$\pm$5.73&  15.29$\pm$4.93&  38.47$\pm$11.72&  72.49$\pm$19.02 \\
  \hline
 Total background                 &  97.64$\pm$14.47 &  66.58$\pm$9.23 &  143.54$\pm$15.65 &  307.76$\pm$35.84 \\
 $\ttbar$ ($\sigma = 6.7$ pb)    &  0.15$\pm$0.03 &  0.18$\pm$0.03 &  0.34$\pm$0.04 &  0.67$\pm$0.06 \\


  \hline
 Observed   & 99 & 96 & 147 & 342 \\
  \hline\hline
                       &          &            &         &      \\
                       &          &            &         &      \\
  \hline\hline
\multicolumn{5}{c}{ N$_{\rm jet}$ = 1 Control Sample per Dilepton Flavor Category}\\
  \hline
 Source               &    $ee$     & $\mu\mu$   &   $e\mu$    &   $\ell\ell$ \\
  \hline

 $WW$   &  9.74$\pm$1.08&  8.73$\pm$0.97&  21.55$\pm$2.31&  40.01$\pm$4.24 \\
 $WZ$   &  4.95$\pm$0.25&  2.66$\pm$0.15&  4.11$\pm$0.21&  11.72$\pm$0.52 \\
 $ZZ$   &  1.59$\pm$1.23&  1.58$\pm$1.22&  0.91$\pm$0.70&  4.08$\pm$3.14 \\
 $W\gamma$ &  3.70$\pm$1.47&  0.00$\pm$0.00&  4.43$\pm$1.12&  8.14$\pm$2.27 \\
 DY$\rightarrow\tau\tau$     &  4.64$\pm$0.83&  4.42$\pm$0.78&  8.81$\pm$1.50&  17.87$\pm$2.99 \\
 DY$\rightarrow ee+\mu\mu$   &  21.04$\pm$4.27&  16.48$\pm$3.56&  3.31$\pm$0.90&  40.83$\pm$7.28 \\
 $W$+jet fakes &  12.14$\pm$3.73&  14.84$\pm$4.60&  67.26$\pm$20.43&  94.23$\pm$26.16 \\
  \hline
 Total background                 &  57.80$\pm$8.97 &  48.70$\pm$7.57 &  110.37$\pm$21.27 &  216.87$\pm$32.46 \\
 $\ttbar$ ($\sigma = 6.7$ pb)    &  3.94$\pm$0.22 &  4.02$\pm$0.22 &  9.47$\pm$0.49 &  17.44$\pm$0.86 \\


  \hline
 Observed   & 58 & 54 & 107 & 219 \\
  \hline\hline
   \end{tabular} 
 \end{center} 
\label{tab:cntrl_bkgr_2fb}
\end{table*}

\section{\label{sec:ttbar} $\ttbar$  using 2 jet selection}

\begin{figure*}[htbp]
 \begin{center}
  \begin{tabular}[t]{cc}
    \includegraphics[width=0.45\textwidth, clip]{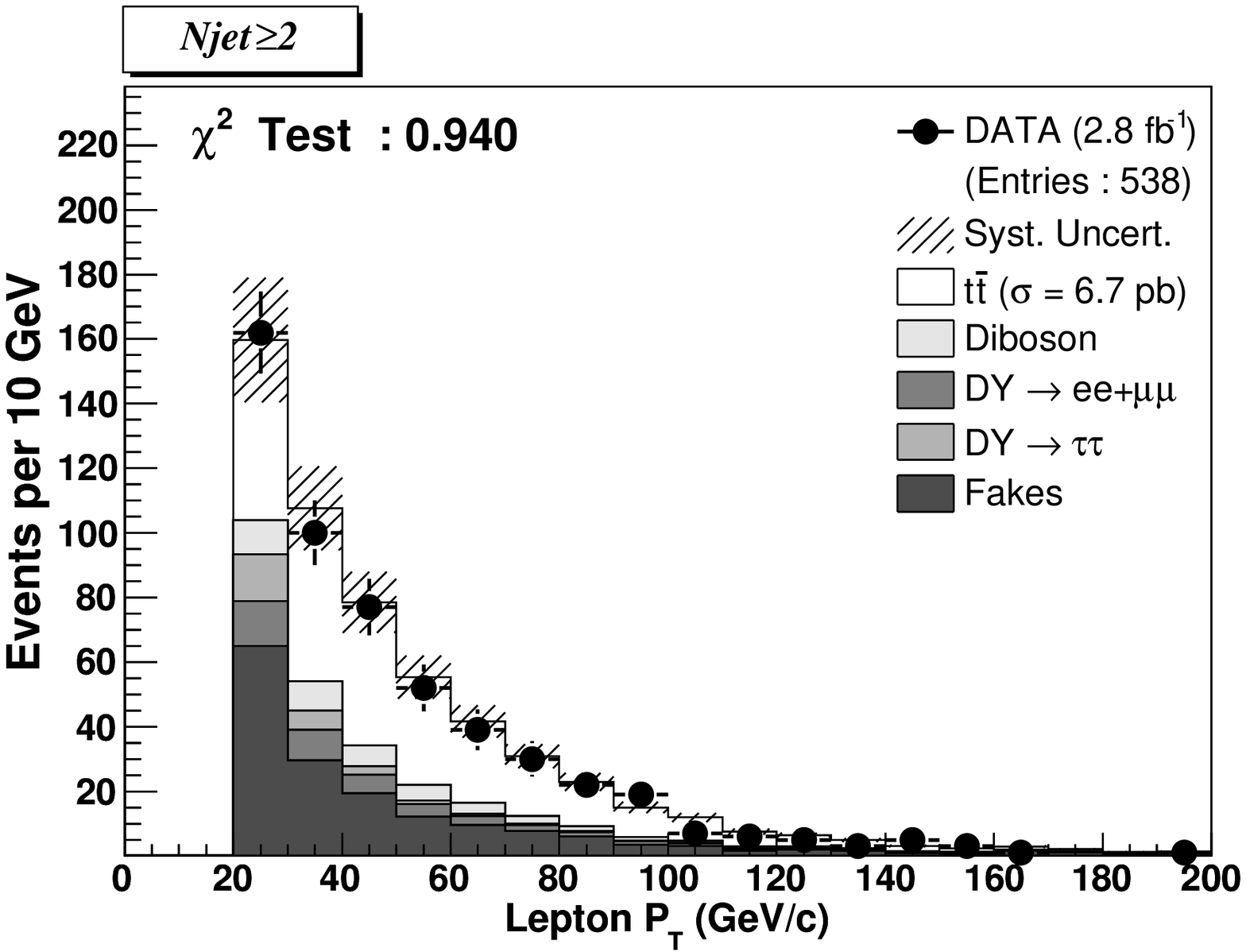} &    
    \includegraphics[width=0.45\textwidth, clip]{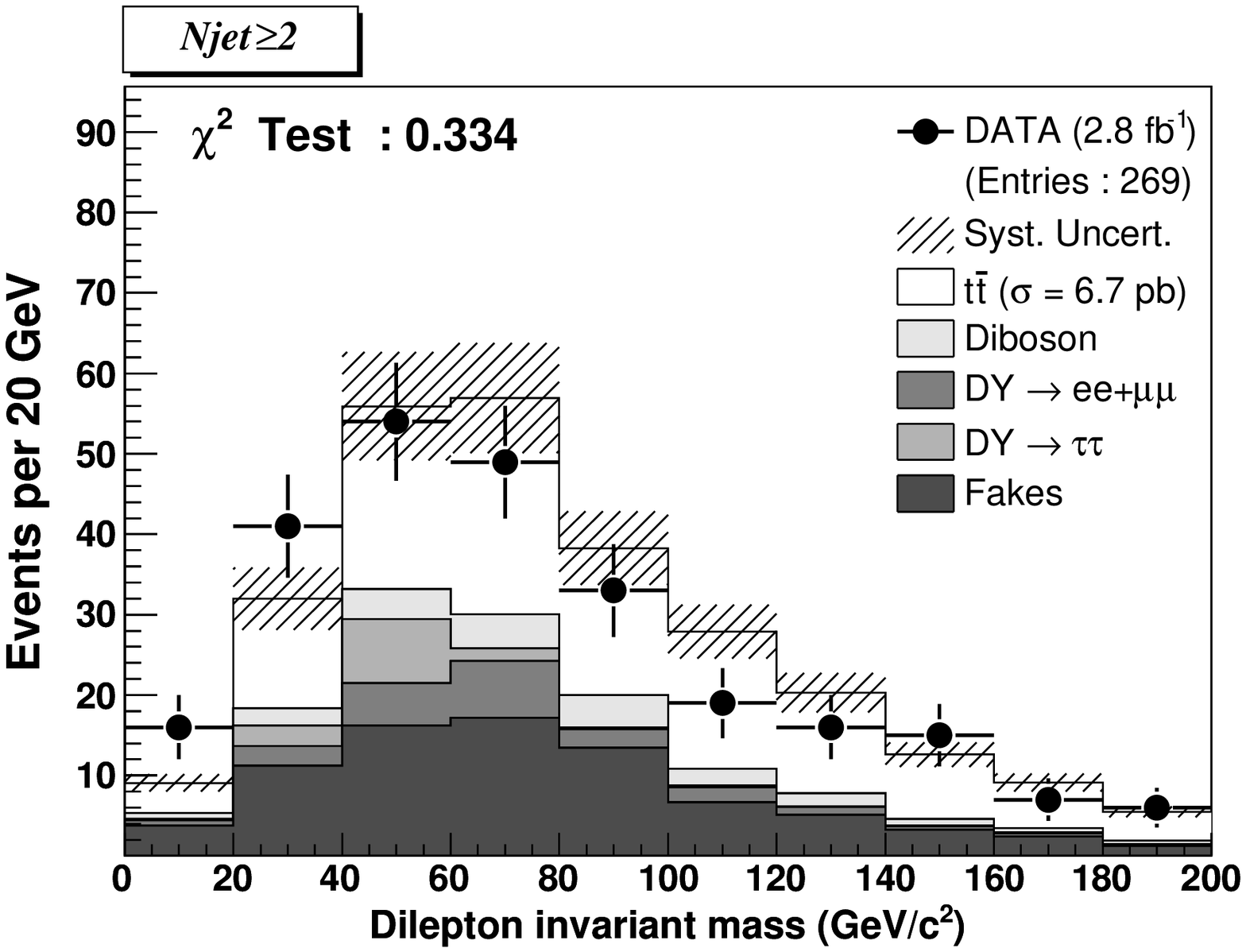} \\  
    \includegraphics[width=0.45\textwidth, clip]{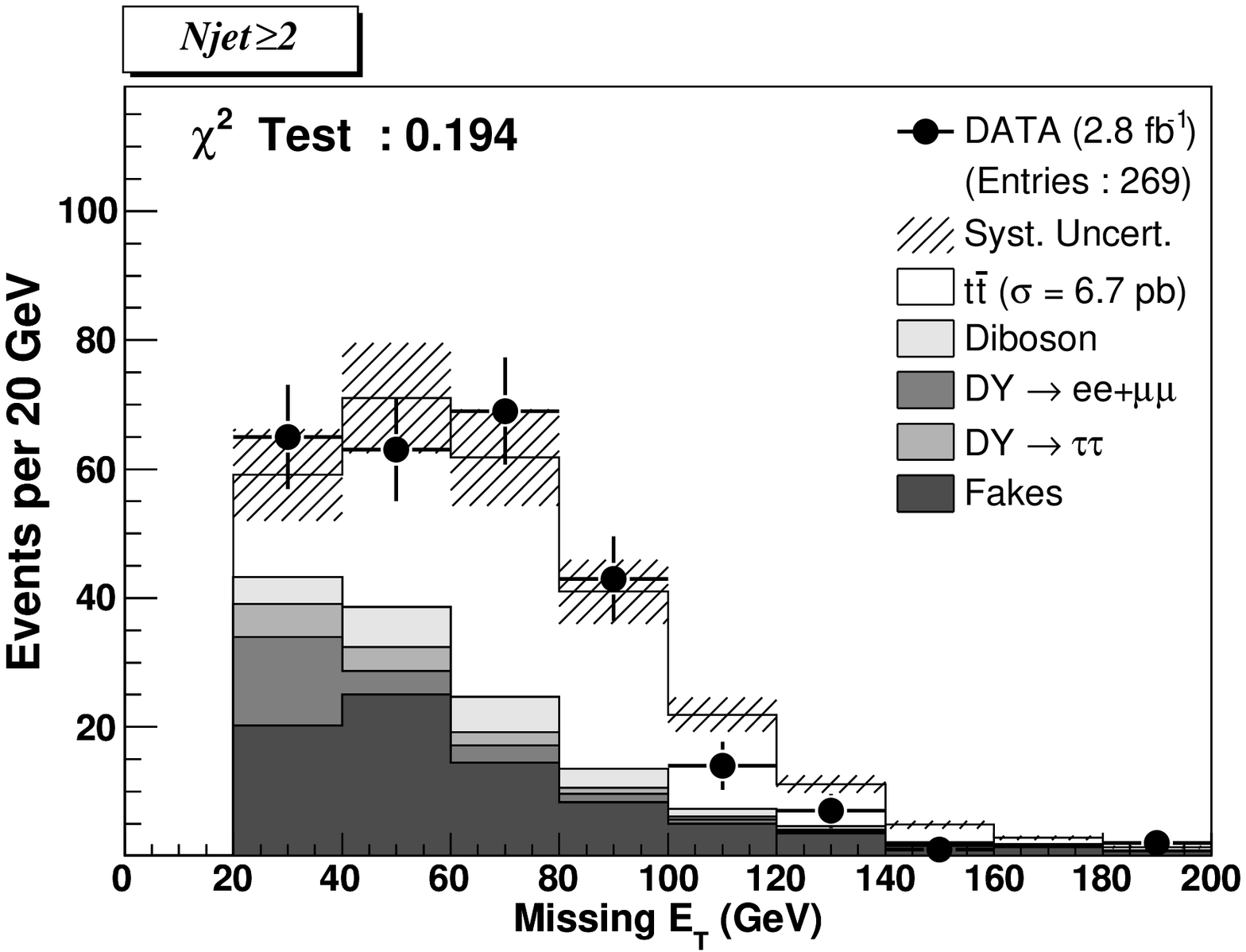} &	    
    \includegraphics[width=0.45\textwidth, clip]{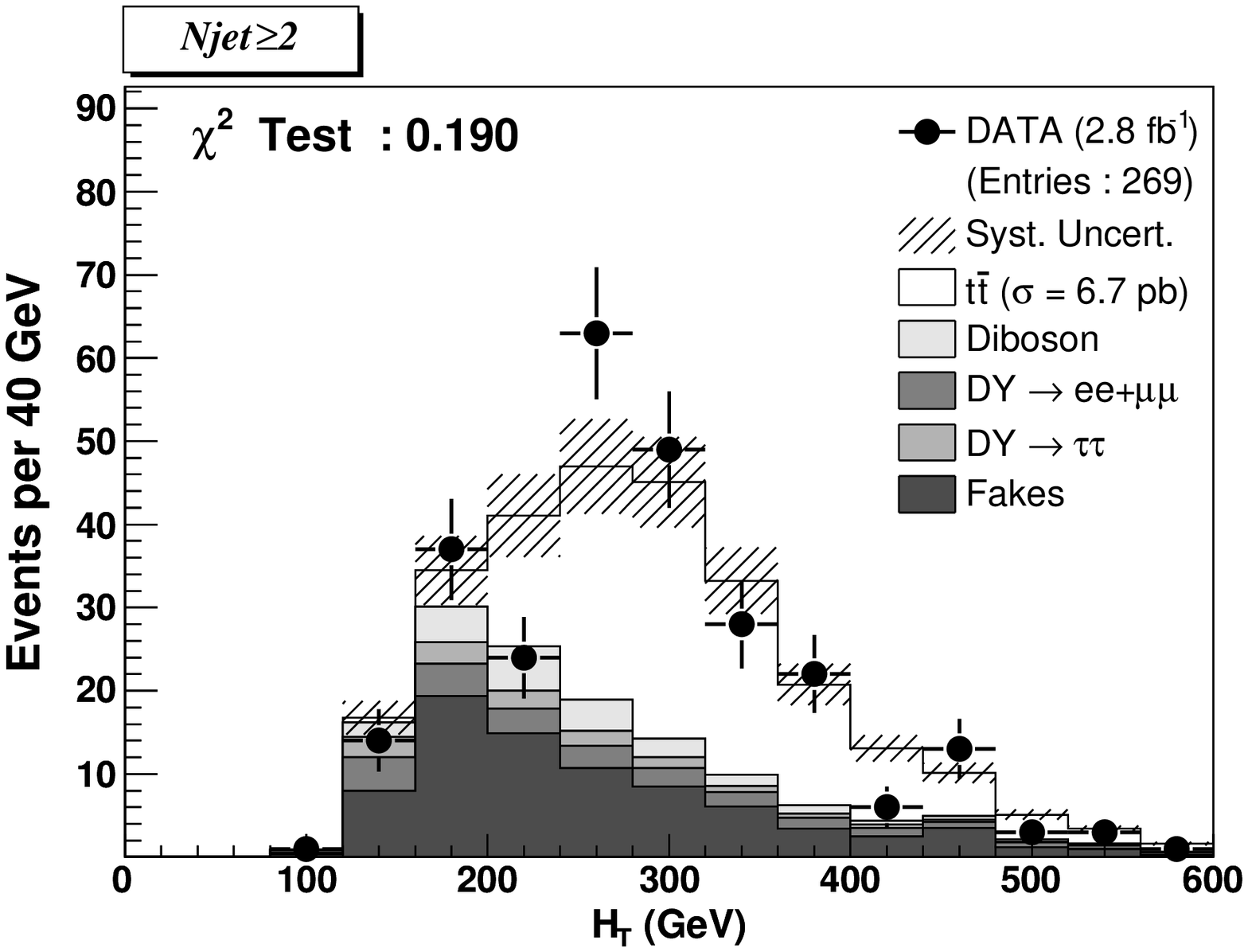} \\         
  \end{tabular}
 \end{center}
  \caption{ Background and top quark signal predictions 
            overlaid on the data for 2-jet events before the H$_T$ and 
            the opposite lepton charge requirement in 2.8~fb$^{-1}$.
            From top left to bottom right: 
            Two leptons transverse energy spectrum,
            the dilepton invariant mass, the event $\met$ and H$_T$. 
            The  hatched area represents the uncertainty in the
            total background estimate.} 
  \label{fig:2jet_beforeHTOS}
\end{figure*}

As an intermediate step toward the final result, 
Table~\ref{tab:2jet_beforeHTOS} shows
the predictions for signal and background in events with
two or more jets passing all of the $\ttbar$ selection criteria, but  
the H$_T > 200$~GeV and the opposite lepton charge requirement.
For this sample the $\ttbar$ signal contribution is almost equal to the total
background contribution.
There is good agreement between predicted and observed number of events
both in overall normalization and in the bin-by-bin distribution 
for the same four kinematic variables used in the 0 and 1-jet control samples,
as shown in Fig.~\ref{fig:2jet_beforeHTOS}.
The agreement is quantified
in terms of the probability for
the $\chi^{2}$/ndf distribution and 
shown as ``$\chi^{2}$ Test'' on the figures.

\begin{table*}
  \caption{ Summary table by lepton flavor content, 
            of background estimates, $\ttbar$ predictions 
            and observed events in data corresponding to 
            an integrated luminosity of 2.8~fb$^{-1}$ for
            the $\ge$2 jet bin before the $\Ht$ and the opposite lepton 
            charge requirement events. The uncertainties
            are the sums in quadrature of the statistical and systematic 
            errors.
            The last column is the total dilepton sample obtained as
            the sum of the $ee$, $\mu\mu$ and $e\mu$ contributions. }
 \begin{center}
  \begin{tabular}{l c c c c}
  \hline\hline
\multicolumn{5}{c}{ N$_{\rm jet} \ge$ 2 $\ttbar$ Sample per Dilepton Flavor Category}\\
  \hline
 Source              &    $ee$     & $\mu\mu$   &   $e\mu$    &   $\ell\ell$ \\
  \hline

 $WW$   &  3.54$\pm$0.63&  3.65$\pm$0.65&  7.50$\pm$1.28&  14.70$\pm$2.47 \\
 $WZ$   &  1.75$\pm$0.23&  1.01$\pm$0.14&  1.68$\pm$0.23&  4.44$\pm$0.57 \\
 $ZZ$   &  0.83$\pm$0.65&  0.74$\pm$0.58&  0.47$\pm$0.37&  2.04$\pm$1.59 \\
 $W\gamma$ &  0.62$\pm$0.41&  0.00$\pm$0.00&  1.45$\pm$0.58&  2.07$\pm$0.78 \\
 DY$\rightarrow\tau\tau$     &  2.97$\pm$0.75&  3.29$\pm$0.84&  6.68$\pm$1.67&  12.94$\pm$3.22 \\
 DY$\rightarrow ee+\mu\mu$   &  20.33$\pm$6.00&  8.04$\pm$2.73&  1.76$\pm$0.72&  30.13$\pm$8.54 \\
 $W$+jet fakes &  9.66$\pm$3.00&  18.67$\pm$5.77&  54.67$\pm$16.59&  83.00$\pm$22.90 \\
  \hline
 Total background                 &  39.71$\pm$8.73 &  35.40$\pm$7.36 &  74.22$\pm$17.13 &  149.33$\pm$28.19 \\
 $\ttbar$ ($\sigma = 6.7$ pb)    &  31.25$\pm$1.52 &  32.69$\pm$1.59 &  74.62$\pm$3.58 &  138.56$\pm$6.61 \\


\hline
 Observed   & 58 & 68 & 143 & 269 \\
  \hline\hline
  \end{tabular} 
\end{center}
  \label{tab:2jet_beforeHTOS}
\end{table*}

\section{\label{sec:Results}Results}

\begin{figure*}[htbp]
 \begin{center}
  \begin{tabular}[t]{cc}
    \includegraphics[width=0.45\textwidth, clip]{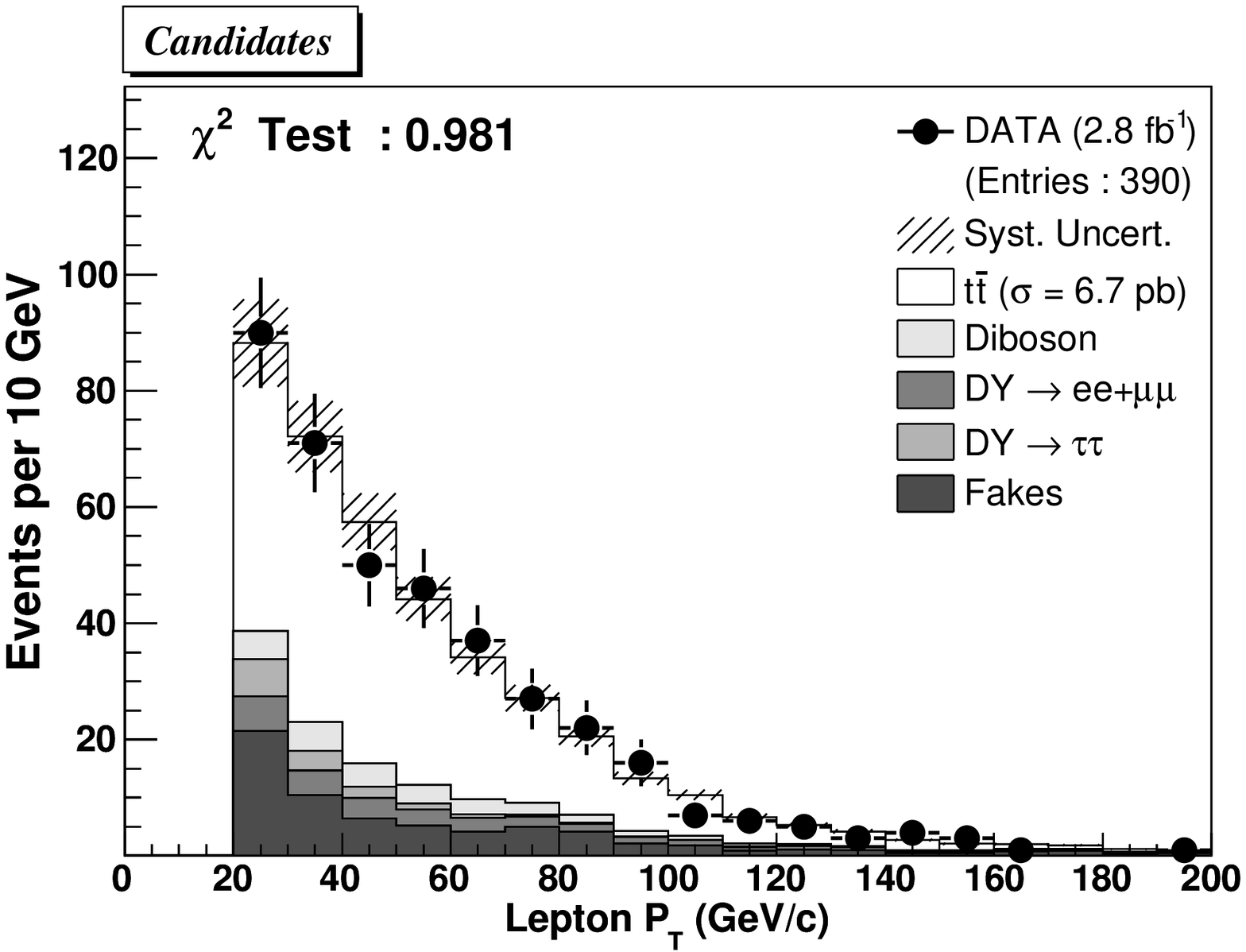} &    
    \includegraphics[width=0.45\textwidth, clip]{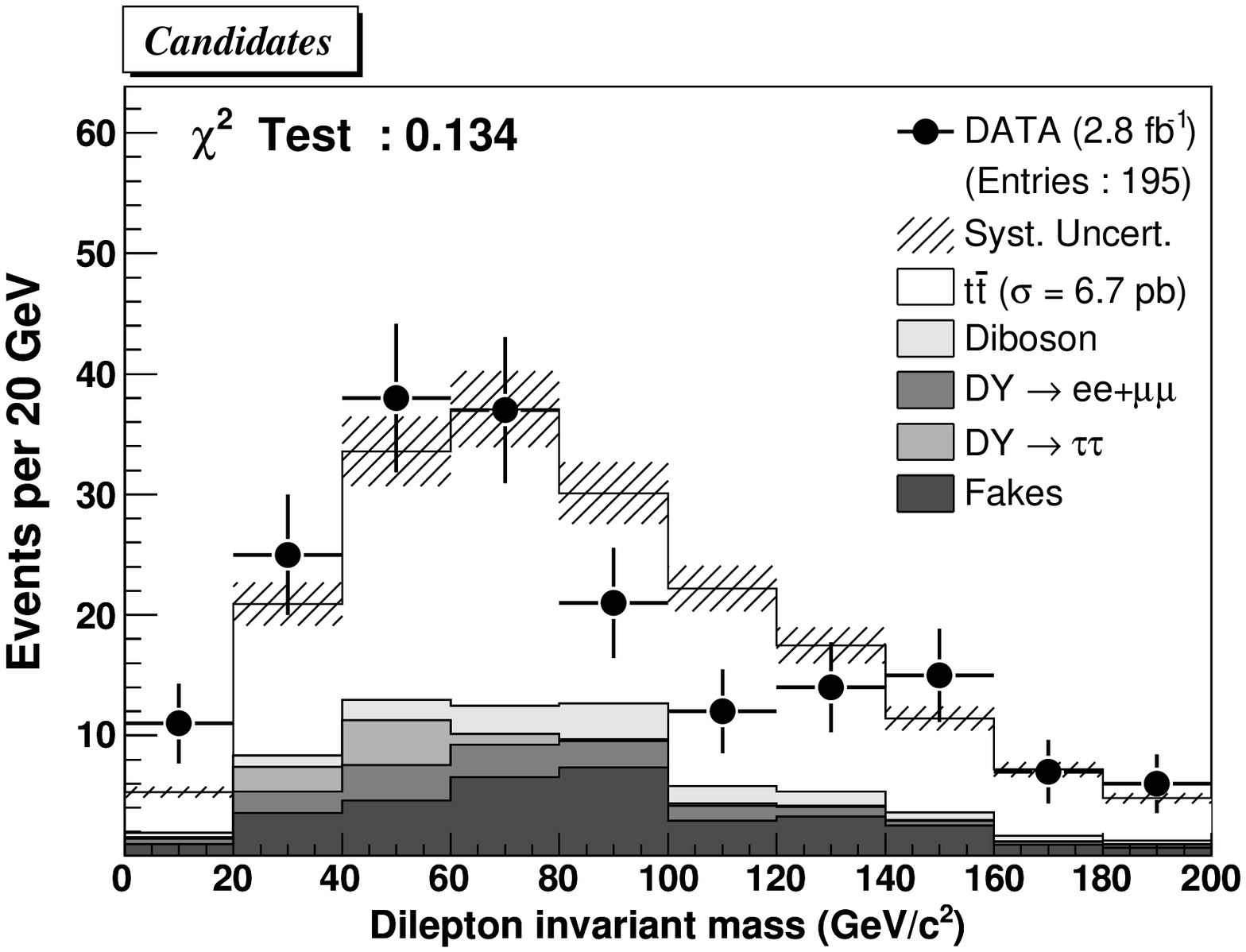} \\  
    \includegraphics[width=0.45\textwidth, clip]{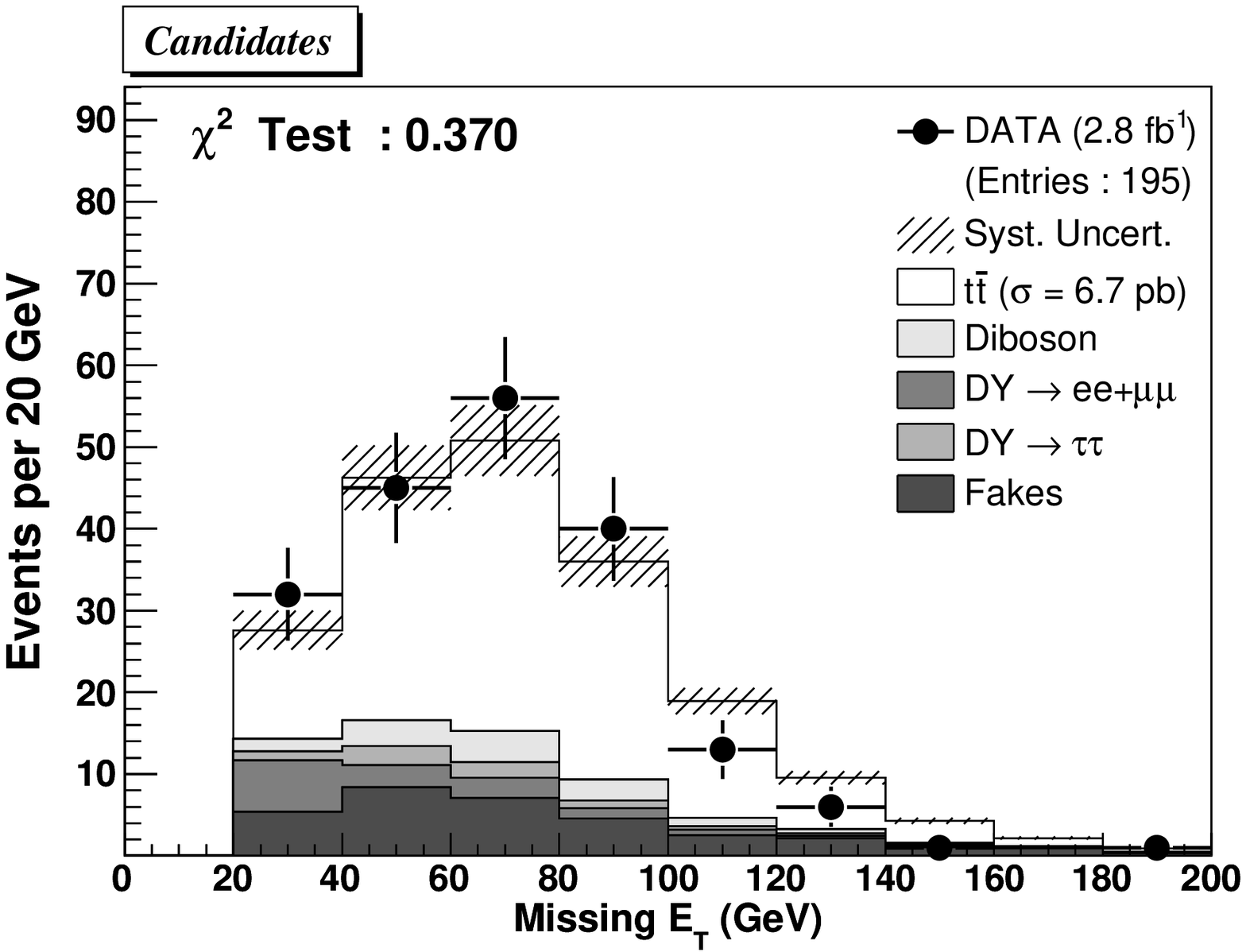} &          
    \includegraphics[width=0.45\textwidth, clip]{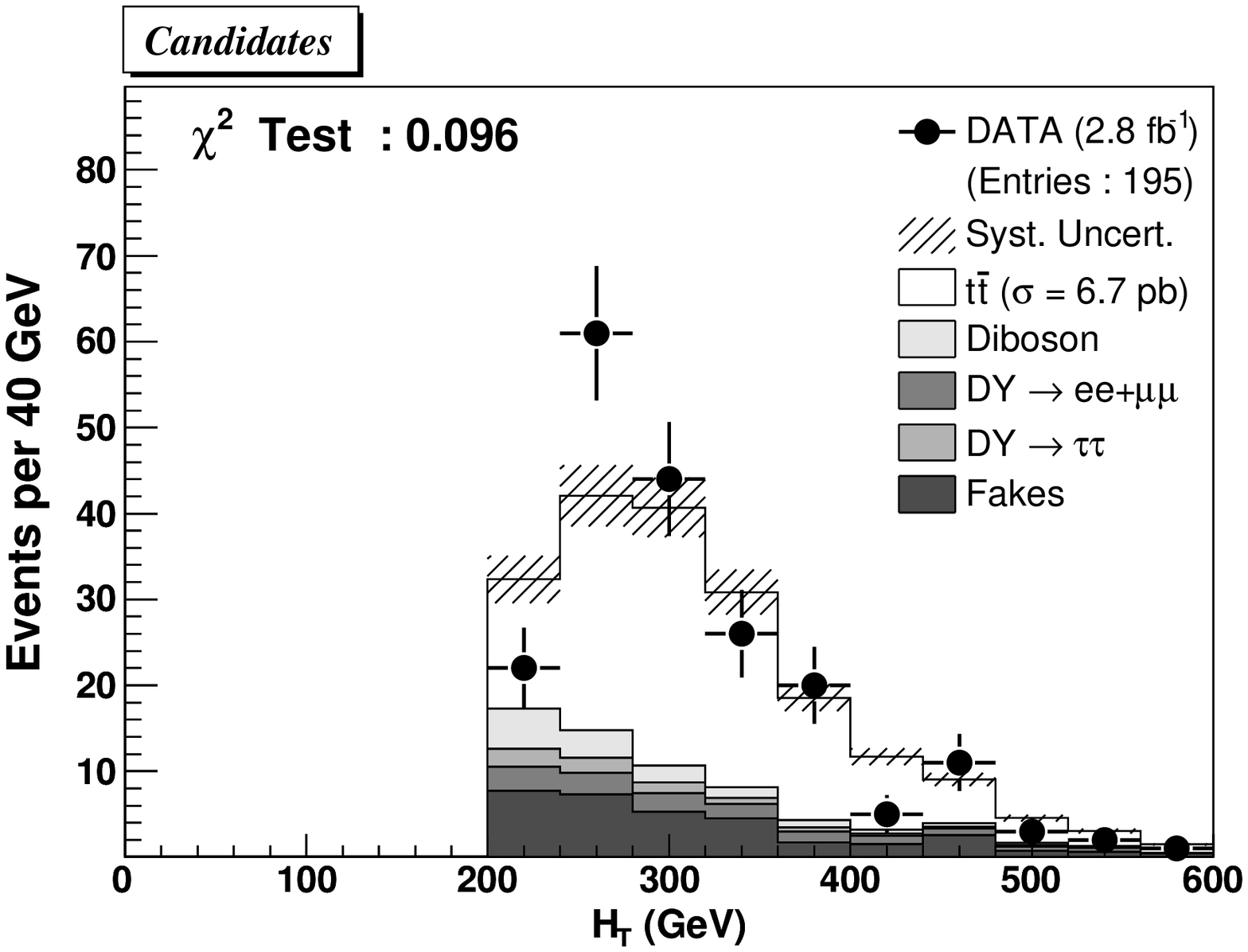} \\         
  \end{tabular}
 \end{center}
  \caption{ Background and top quark signal predictions 
            overlaid on the data for top quark DIL candidate events in 2.8~fb$^{-1}$.             From top left to bottom right: 
            Two leptons transverse energy spectrum,
            the dilepton invariant mass, the event $\met$ and H$_T$. 
            The  hatched area represents the uncertainty in the
            total background estimate.} 
  \label{fig:KinPlots}
\end{figure*}

The signal and background DIL candidate
events, that is events in the 2 or more jet samples after
the final $H_T > 200$~GeV and opposite charge requirements,
are shown in Table~\ref{tab:summary-all-cuts} separately for the different
dilepton flavor contribution. 
The $\ttbar$ rate is computed assuming a 
$\ttbar$ production cross section 
in agreement with the NLO standard model calculation 
for a top mass of 175 GeV/$c^{2}$, of 6.7$^{+0.7}_{-0.9}$~pb
\cite{theory}.
The sum of the background and signal contributions is
labelled ``Total SM expectation'' and can be compared to 
the number of ``Observed'' data in 2.8~fb$^{-1}$. 
Fig.~\ref{fig:KinPlots} shows the $\ttbar$ and the different backgrounds
overlaid on the data, 
for the single lepton $p_T$, dilepton invariant mass, 
the event $\met$ and $\Ht$ distributions.
Again there is overall good agreement between data and total background plus
$\ttbar$ expectations, as shown by the probability for
the $\chi^{2}$/ndf distribution reported on the figures.

\begin{table*}[h]
\caption{Summary table by lepton flavor content of background
         estimates, $\ttbar$ predictions and 
observed events in the final sample of events with $\geq$ 2 jets
passing all candidate selection criteria,
       for an integrated luminosity of 2.8~fb$^{-1}$. The uncertainties
         are the sums in quadrature of the statistical and systematic errors.
         The last column is the total dilepton sample obtained as
         the sum of the $ee$, $\mu\mu$ and $e\mu$ contributions. 
         }
 \begin{center}
 \begin{tabular}{l c c c c}
  \hline\hline
\multicolumn{5}{c}{ $\ttbar$ Signal Events per Dilepton Flavor Category}\\
  \hline
 Source                      &    $ee$        &    $\mu\mu$    &   $e\mu$       &   $\ell\ell$ \\
  \hline

 $WW$   &  2.16$\pm$0.38&  2.42$\pm$0.42&  4.79$\pm$0.80&  9.37$\pm$1.51 \\
 $WZ$   &  0.94$\pm$0.15&  0.68$\pm$0.11&  0.59$\pm$0.10&  2.22$\pm$0.33 \\
 $ZZ$   &  0.65$\pm$0.51&  0.64$\pm$0.50&  0.23$\pm$0.18&  1.51$\pm$1.18 \\
 $W\gamma$ &  0.23$\pm$0.25&  0.00$\pm$0.00&  0.00$\pm$0.00&  0.23$\pm$0.25 \\
 DY$\rightarrow\tau\tau$     &  1.67$\pm$0.32&  1.76$\pm$0.34&  3.87$\pm$0.72&  7.29$\pm$1.34 \\
 DY$\rightarrow ee+\mu\mu$   &  11.81$\pm$2.16&  5.32$\pm$1.23&  1.36$\pm$0.60&  18.49$\pm$2.73 \\
 $W$+jet fakes &  3.91$\pm$1.28&  9.34$\pm$3.05&  20.90$\pm$6.43&  34.15$\pm$9.51 \\
  \hline
 Total background                 &  21.37$\pm$3.14 &  20.16$\pm$3.64 &  31.73$\pm$6.78 &  73.26$\pm$11.30 \\
 $\ttbar$ ($\sigma = 6.7$ pb)    &  28.80$\pm$1.41 &  31.24$\pm$1.52 &  70.15$\pm$3.36 &  130.19$\pm$6.21 \\
  \hline                      
 Total SM expectation &  50.17$\pm$4.25 &  51.40$\pm$5.00 &  101.88$\pm$10.06 &  203.45$\pm$17.33 \\

  \hline
 Observed             & 41              & 55              & 99              & 195 \\
  \hline\hline
  \end{tabular}
 \end{center}
 \label{tab:summary-all-cuts}
\end{table*}

\par

\begin{figure*}[h]
\begin{center}
  \includegraphics[width=10cm]{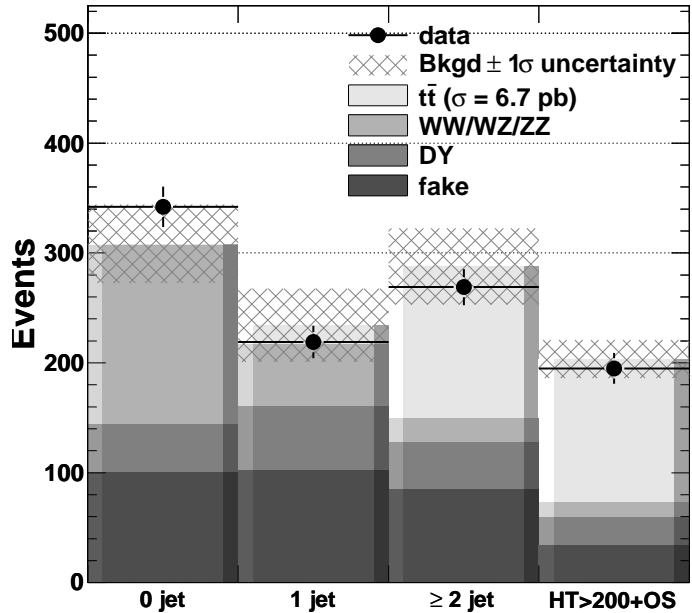} 
  \caption{ Dilepton candidate events (black point) by
           jet multiplicity. The stacked histogram represents the background
           contribution and the $\ttbar$ contribution 
           for an assumed $ \sigma_{\ttbar}$ = 6.7~pb. 
           The hatched area is the uncertainty in the total
           background estimate.}
\label{fig:jet_multiplicities}
\end{center}
\end{figure*}

Table~\ref{tab:summary-njet} summarizes the background and signal
predictions for the 0, 1 and $\ge 2$ jet-bin control samples and for the
signal sample. 
Figure \ref{fig:jet_multiplicities} shows the overall number of candidate events 
in the different jet multiplicity bins overlaid on top of a stacked histogram 
of the different background components.  
The  band gives the $\ttbar$
contribution for a cross section of 6.7~pb. The 
hatched area represents the uncertainty in the total background estimate.
From the difference between the observed data and the total background
predictions, we measure a cross section 
for $\ttbar$ events in the dilepton channel of: 
\begin{center}
$\sigma_{\ttbar}$ = 6.27 $\pm$ 0.73(stat) $\pm$ 0.63(syst) $\pm$ 0.39(lum) pb.
\end{center}
where the first uncertainty is statistical, the second is the
convolution of the acceptance and background systematics and the
third comes from the 6\% uncertainty in the luminosity
measurement. This results assumes a top quark mass of M$_{t}$ = 175 GeV/$c^2$.
Studies of the DIL selection acceptance vs M$_t$ shows 
an increase in acceptance of 3\% for each 1 GeV/$c^2$, 
in the range $\pm$ 2~GeV/$c^2$ 
around the combined Tevatron top quark mass measurement of
M$_{t} = 173.1 \pm 0.6_{\rm stat} \pm 1.1_{\rm syst}$~GeV/$c^2$~\cite{mtop}.
The theory cross section decreases by approximately 0.2 pb for each 1 GeV/$c^{2}$
increase over the mass range from 170 to 180 GeV/$c^{2}$.

\begin{table*}[h]
 \caption{Summary table by jet multiplicity bin 
          of background estimates, $\ttbar$ predictions 
          and observed events in data corresponding to 
          an integrated luminosity of 2.8~fb$^{-1}$. 
          The uncertainties are the sums in quadrature
          of the statistical and systematic error.
          The last column contains the candidate
          events with $\Ht > 200$~GeV and opposite sign 
          lepton cuts applied.} 
 \begin{center}
  \begin{tabular}{l c c c c}
  \hline\hline
\multicolumn{5}{c}{Control Sample and Signal Events per Jet Multiplicity} \\
  \hline
 Source &   0 jet    &    1 jet      &  $\ge 2$ jet   &  $\Ht$ + OS \\
  \hline

 $WW$                           &  142.87$\pm$12.57&    40.01$\pm$4.24&    14.70$\pm$2.47&    9.37$\pm$1.51 \\  
 $WZ$                           &   11.65$\pm$0.74&     11.72$\pm$0.52&     4.44$\pm$0.57&    2.22$\pm$0.33 \\  
 $ZZ$                           &    8.79$\pm$6.78&      4.08$\pm$3.14&     2.04$\pm$1.59&    1.51$\pm$1.18 \\  
 $W\gamma$                      &   28.03$\pm$7.11&      8.14$\pm$2.27&     2.07$\pm$0.78&    0.23$\pm$0.25 \\  
 DY$\to\tau\tau$                &    3.89$\pm$0.58&     17.87$\pm$2.99&    12.94$\pm$3.22&    7.29$\pm$1.34 \\  
 DY$\to$ $ee+\mu\mu$            &   40.04$\pm$5.77&     40.83$\pm$7.28&    30.13$\pm$8.54&   18.49$\pm$2.73 \\  
 $W$+jet fakes                  &   72.49$\pm$19.02&    94.23$\pm$26.16&   83.00$\pm$22.90&  34.15$\pm$9.51 \\  
  \hline                                                                                                        
 Total background               &  307.76$\pm$35.84 &  216.87$\pm$32.46 & 149.33$\pm$28.19&  73.26$\pm$11.30 \\ 
 $\ttbar$ ($\sigma = 6.7$ pb)   &    0.67$\pm$0.06 &    17.44$\pm$0.86 &  138.56$\pm$6.61&  130.19$\pm$6.21 \\  
  \hline                                                                                                        
Total SM expectation            &  308.43$\pm$35.87 &  234.31$\pm$33.28 & 287.89$\pm$34.70& 203.45$\pm$17.33 \\ 

 \hline
 Observed                       & 342 & 219 & 269 & 195 \\
  \hline\hline
  \end{tabular}
 \end{center}
\label{tab:summary-njet}
\end{table*}

%

\par 
As a test of lepton universality, 
we quote the results for the individual dilepton flavor decay modes:
\begin{center}
$\sigma_{\ttbar}(ee)$ = 4.57 $\pm$ 1.56(stat) $\pm$ 0.58(syst)  pb  \\
$\sigma_{\ttbar}(\mu\mu)$ = 7.47 $\pm$ 1.63(stat) $\pm$ 0.79(syst) pb  \\
$\sigma_{\ttbar}(e\mu)$ = 6.43 $\pm$ 0.95(stat) $\pm$ 0.69(syst)  pb. 
\end{center}
All of the results are consistent with each other. 
Similar conclusions hold for the cross section of signal events 
where both leptons are isolated, which is a sample extensively used in other
SM precision measurements like the $Z$ cross section measurement:
\begin{center}
$\sigma_{\ttbar}$(iso) = 6.40 $\pm$ 0.75(stat) $\pm$ 0.49(syst) pb.
\end{center}
The luminosity error is common 
to all of these subsamples and is not explicitely quoted.

\section{\label{sec:Concl}Conclusions}
We present a measurement of the $\ttbar$ cross section at the Tevatron
in a sample of data corresponding to an integrated luminosity of 2.8~fb$^{-1}$ 
collected by the CDF II detector.
Using events with two leptons, large missing energy
and two or more jets we select a sample with a signal
top quark pairs almost a factor of two larger than the background.
The contamination from SM sources is checked in lower jet multiplicity samples.
From the excess of data over the background predictions we measure:
\begin{center}
$\sigma_{\ttbar}$ = 6.27 $\pm$ 0.73(stat) $\pm$ 0.63(syst) $\pm$ 0.39(lum) pb,
\end{center}
\noindent
or
\begin{center}
$\sigma_{\ttbar}$ = 6.27 $\pm$ 1.03(total)  pb,
\end{center}
\noindent
consistent with the NLO standard model calculation of 6.7$^{+0.7}_{-0.9}$~pb.
Yields in the $ee$, $\mu\mu$ and $e\mu$ final states are in agreement with
the predictions from lepton universality.

\section{Acknowledgments}
We thank the Fermilab staff and the technical staffs 
of the participating institutions for their vital contributions. 
This work was supported by the U.S. Department of Energy and National Science Foundation; 
the Italian Istituto Nazionale di Fisica Nucleare; the Ministry of Education, Culture, Sports, 
Science and Technology of Japan; the Natural Sciences and Engineering Research Council of Canada; 
the National Science Council of the Republic of China; the Swiss National Science Foundation; 
the A.P. Sloan Foundation; the Bundesministerium f\"ur Bildung und Forschung, Germany; 
the World Class University Program, the National Research Foundation of Korea; 
the Science and Technology Facilities Council and the Royal Society, UK; 
the Institut National de Physique Nucleaire et Physique des Particules/CNRS; 
the Russian Foundation for Basic Research; the Ministerio de Ciencia e Innovaci\'{o}n, 
and Programa Consolider-Ingenio 2010, Spain; the Slovak R\&D Agency; and the Academy of Finland. 

\end{document}